\documentclass{article}

\usepackage{PRIMEarxiv}

\usepackage[utf8]{inputenc} 
\usepackage[T1]{fontenc}    
\usepackage{hyperref}       
\usepackage{url}            
\usepackage{booktabs}       
\usepackage{amsfonts}       
\usepackage{nicefrac}       
\usepackage{microtype}      
\usepackage{xcolor}         
\usepackage{times}
\usepackage{soul}

\usepackage{balance} 
\usepackage{algorithm}
\usepackage{algpseudocode}
\usepackage{enumitem}
\usepackage{graphicx}
\usepackage{subcaption}   
\usepackage{float}
\usepackage{mathtools}

\usepackage{amsmath}
\usepackage{amsthm}

\theoremstyle{plain}
\newtheorem{theorem}{Theorem}

\theoremstyle{definition}
\newtheorem{definition}[theorem]{Definition}

\theoremstyle{remark}

\title{Learning Communication Skills in Multi-task Multi-agent Deep Reinforcement Learning
}

\author{
  Changxi Zhu \\
  Department of Information and Computing Sciences \\
  Utrecht University \\
  Utrecht, Netherlands\\
  \texttt{c.zhu@uu.nl} \\
   \And
  Mehdi Dastani \\
  Department of Information and Computing Sciences \\
  Utrecht University \\
  Utrecht, Netherlands\\
  \texttt{m.m.dastani@uu.nl} \\
  \AND
  Shihan Wang \\
  Department of Information and Computing Sciences \\
  Utrecht University \\
  Utrecht, Netherlands\\
  \texttt{s.wang2@uu.nl} \\
}

\begin{document}
\maketitle

\begin{abstract}
In multi-agent deep reinforcement learning (MADRL), agents can communicate with one another to perform a task in a coordinated manner. When multiple tasks are involved, agents can also leverage knowledge from one task to improve learning in other tasks. In this paper, we propose Multi-task Communication Skills (MCS), a MADRL with communication method that learns and performs multiple tasks simultaneously, with agents interacting through learnable communication protocols. MCS employs a Transformer encoder to encode task-specific observations into a shared message space, capturing shared communication skills among agents. To enhance coordination among agents, we introduce a prediction network that correlates messages with the actions of sender agents in each task. We adapt three multi-agent benchmark environments to multi-task settings, where the number of agents as well as the observation and action spaces vary across tasks. Experimental results demonstrate that MCS achieves better performance than multi-task MADRL baselines without communication, as well as single-task MADRL baselines with and without communication.
\end{abstract}


\section{Introduction}

Communication is essential in multi-agent deep reinforcement learning (MADRL) for enabling information sharing and enhancing collaboration among multiple agents toward common goals \cite{Changxi2024Survey}, especially in partially observable environments such as autonomous driving \cite{Shai2016MARLAD}, sensor networks \cite{Pipattanasomporn2009Smart}, and multi-robot control \cite{Kober2013Robotics}. Recent works on learning communication in MADRL have investigated what information to communicate \cite{Jiang2018ATOC,Yuan2022MAIC}, when and with whom to communicate \cite{Singh2019IC3Net,Ding2020I2C,Zhang2025TGCNet} and how to integrate communication into policies \cite{Das2019TarMAC,Guan2022MASIA}, thereby enabling flexible and dynamic communication. However, these approaches are limited to single-task settings, where both policies and communication protocols are designed and evaluated for only one task. 

A different line of work in the MADRL literature focuses on multi-task learning, which aims to generalize policies across multiple tasks \cite{Zhang2022Survey}. This is achieved either via task-invariant architectures \cite{Hu2021UPDet,Tian2024DT2GS} that exploit the shared structure across tasks to learn a single unified model, or via task representation learning \cite{Qin2024MATTAR,Mahajan2024Embedding} that explicitly models environment dynamics to generalize to unseen tasks. However, despite these achievements, previous works on multi-task learning in MADRL has not considered communication among agents within tasks. In contrast, humans can leverage knowledge, in particular communication knowledge, to learn multiple tasks simultaneously. 


In this work, we extend the scope of MADRL with communication approaches from single-task to multi-task settings. 
The integration of communication into multi-task MADRL is non-trivial as it involves complex processes regarding what, when, and how to communicate. Moreover, tasks may differ in observation and action spaces, making message generation challenging. To address these challenges, we propose encoding task-specific observations into a shared message space, capturing common communication skills that can generalizes across tasks, therefore enabling effective and efficient communication in Multi-task Multi-Agent Deep Reinforcement Learning (Multi-task MADRL). 

In this paper, we propose Multi-task Communication Skills (MCS), a multi-task MADRL with communication method. MCS leverages task-invariant architectures to learn shared communication protocols across tasks, thereby improving coordination among agents and accelerate learning performance in multi-task settings. During communication in each task, messages are encoded by Transformer encoders and exchanged over a learned communication graph, where unnecessary messages are pruned. The received messages are then aggregated through a shared aggregation function and used to attend to important features in policy and critic networks. To further enhance coordination, we introduce a prediction network that maximizes mutual information between messages and actions to encourage informative communication. To evaluate MCS, we conduct experiments on the well-known SMAC benchmark \cite{Samvelyan2019StarCraft}, a novel multi-task AliceBob environment, and an adapted multi-task version of Google Research Football \cite{Kurach2020Football}. We evaluate both the average performance across tasks and the per-task performance. Our results show that MCS significantly outperforms multi-task MADRL baselines without communication, as well as single-task MADRL baselines both with and without communication. We further conduct ablation studies to assess the effects of pruning unnecessary messages and of the prediction network on learning performance. We also analyze the sensitivity of hyperparameters and the learned representations of messages. As a result, agents in MCS can communicate and perform efficiently and effectively across multiple tasks with varying numbers of agents and diverse observation and action spaces.


\section{Related Work}

\paragraph{\textbf{Multi-task Learning in MADRL}} Early approaches in multi-task MADRL, such as DEC-HDRQN \cite{Omidshafie2017Multitask}, distilled single-task Q-networks into a unified Q-network through a distillation process. More recent methods focus either on task-invariant architectures \cite{Iqbal2021REFIL,Hu2021UPDet,Tian2024DT2GS,Li2025RIT} or on learning task representations \cite{Schafer2023Learning,Qin2024MATTAR,Mahajan2024Embedding}. multi-task MADRL methods based on task-invariant architectures exploit shared task structure by unifying inputs across tasks using entity-based representations of observations and actions \cite{Christian2019Common}. Specifically, REFIL \cite{Iqbal2021REFIL} partitions agents into subgroups to capture shared local coordination patterns across tasks. UPDet \cite{Hu2021UPDet} leverages a single unified Transformer architecture to handle varying entities across tasks. DT2GS \cite{Tian2024DT2GS} decomposes each task into multiple sub-tasks via latent variables generated by Transformer encoders. RIT \cite{Li2025RIT} applies mask techniques to selectively disable network layers corresponding to invalid entities. On the other hand, task representations can be learned from a set of tasks based on trajectories \cite{Schafer2023Learning}, transition and reward functions \cite{Qin2024MATTAR}, or task similarity measures \cite{Mahajan2024Embedding}. 

The above research works focus on online multi-task MADRL. A related line of research \cite{Zhang2023ODIS, Chen2024Variational, Liu2025HiSSD} investigates offline multi-task MADRL which utilizes offline data across multiple independent tasks to learn generalized policies for unseen tasks. Among them, ODIS \cite{Zhang2023ODIS} identifies coordination skills for individual agents from states and joint actions in the offline data. HiSSD \cite{Liu2025HiSSD} adopts a hierarchical policy that jointly learns common and task-specific skills in an offline dataset. However, we consider online MADRL without assuming the availability of offline data or state information. Moreover, existing literature on both online and offline multi-task MADRL does not model communication and interactions among agents within tasks, which our work explicitly addresses.

\paragraph{\textbf{Communication in MADRL}} 
Existing research on communication in MADRL primarily focuses on the single-task setting. While our method is the first work on multi-task setting and utilizes a different communication method compared to the existing literature, we get inspiration from single-task MADRL with communication from two main perspectives: what to communicate and when to communicate. First, the content of messages (what to communicate) is generally encoded information derived from either observations \cite{Sukhbaatar16CommNet, Singh2019IC3Net, Wang2020NDQ, Guan2022MASIA, Zhang2025TGCNet} or intended actions \cite{Jiang2018ATOC, Kim2021IS, Yuan2022MAIC}. Specifically, TGCNet \cite{Zhang2025TGCNet} leverages a Transformer encoder to encode local observations as messages. ATOC \cite{Jiang2018ATOC} learns an intention model that encodes both local observations and action intentions. IS \cite{Kim2021IS} encodes imagined trajectories capturing agents’ future action plans. MAIC \cite{Yuan2022MAIC} uses a teammate model to predict teammates’ intentions and generate agent-specific messages. These approaches use an intention model during both training and execution. In contrast, our method learns messages that correlate with actions using a prediction network, which is used only in training. 

The decision of whether to communicate or not can be determined based on confidence or distance measures \cite{Zhang2019VBC,Zhang2020TMC,Han2023MBC}, a binary classifier \cite{Singh2019IC3Net,Mao2020GatedACML,Ding2020I2C,Sun2024T2MAC}, or a communication graph \cite{Das2019TarMAC,Liu2020G2ANet,Jiang2020DGN,Niu2021MAGIC,Hu2024CommFormer,Zhang2025TGCNet}. Specifically, agents can communicate when their confidence is low, as in VBC \cite{Zhang2019VBC}, or choose not to communicate when the previous or estimated messages are similar to the current messages, as in TMC \cite{Zhang2020TMC} and MBC \cite{Han2023MBC}. Methods based on learnable binary classifiers, including IC3Net \cite{Singh2019IC3Net}, GACML \cite{Mao2020GatedACML}, I2C \cite{Ding2020I2C}, and T2MAC \cite{Sun2024T2MAC}, assign communication labels using a threshold during training. Most similar to our work, attention mechanisms are often employed to construct communication graphs, enabling dynamic and flexible communication. Specifically, TarMAC \cite{Das2019TarMAC} formulates communication as an attention-based message aggregation process. G2ANet \cite{Liu2020G2ANet} employs a two-layer attention mechanism for selective message aggregation. DGN \cite{Jiang2020DGN} uses multi-head attention as a convolution kernel to aggregate information among neighboring agents. MAGIC \cite{Niu2021MAGIC} introduces a hard attention to dynamically construct communication graphs, and CommFormer \cite{Hu2024CommFormer} leverages attention to allocate credit to received messages within a bi-level optimization framework. More recently, TGCNet \cite{Zhang2025TGCNet} learns a dynamic and directed communication graph using a multi-key gated mechanism with multiple hard attention modules. In contrast, our method also employs attention mechanisms but further integrates a threshold to prune unnecessary messages.


\section{Preliminaries}

We extend the definition of multi-task MADRL \cite{Omidshafie2017Multitask} by incorporating entities and communication.

\begin{definition} 
An Entity-Based Multi-Task Multi-Agent Reinforcement Learning with communication problem $\mathcal{MT}$ is defined as:
$$
\mathcal{MT} := \{\mathcal{T}_k | k=1,2,...,K\} \, ,\mathcal{T}_k := \langle I, E, S, A, O, M, \Omega, P, R, \gamma \rangle
$$
where each task $\mathcal{T}_k$ is a Dec-POMDP tuple augmented with an additional message set $M$ and entity set $E$. The Dec-POMDP components are a set of agents $I$, a set of environment states $S$, a set of joint actions $A$, a set of joint observations $O$, an observation function $\Omega$, a transition function $P$, a reward function $R$, and a discount factor $\gamma$. We assume that agents are a subset of entities, i.e., $I \subseteq E$, and states are represented as an entity-based matrix, i.e., $S \subseteq \mathbb{R}^{|E| \times D^e}$, where $D^e$ denotes the feature dimensionality, which remains the same for each entity in $E$. The observation function $\Omega$ maps the entity set to the set of observable entities, $\Omega: E \rightarrow \hat{E}$, which is used to construct observations $O \subseteq \mathbb{R}^{|\hat{E}| \times D^e}$ for $\hat{E} \subseteq E$. To enable a shared policy across tasks, we unify the varying sets of agents, entities, states, actions, observations, and messages by introducing the union sets of agents $\mathcal{I}$, entities $\mathcal{E}$, states $\mathcal{S}$, observations $\mathcal{O}$, actions $\mathcal{A}$, and messages $\mathcal{M}$ across tasks. Since each individual task can have its own state space, we use $S^k \subseteq \mathcal{S}$ to denote the state space of task $k$. We follow the same conventional notation for all components of the Dec-POMDP of task $k$, where we have $I^k \subseteq \mathcal{I}$, $E^k \subseteq \mathcal{E}$, $\hat{E}^k \subseteq \mathcal{E}$, $S^k \subseteq \mathcal{S}$, $O^k \subseteq \mathcal{O}$, $A^k \subseteq \mathcal{A}$, and $M^k \subseteq \mathcal{M}$.
\end{definition}

In the partial observable environment of task $k$, agents may not observe the state $s^k \in S^k$ of the environment. Each agent $i$ receives a local observation $o^k_i \in O^k$ and encodes it into a message $m^k_i \sim f(o^k_i;\theta_{enc})$ parametrized by $\theta_{enc}$. A joint policy parametrized by $\theta_{\pi}$ is then defined as $\pi(\boldsymbol{a}^k \mid \boldsymbol{o}^k, \boldsymbol{m}^k;\theta_{\pi})$, where joint observations $\boldsymbol{o}^k = \langle o^k_i,...,o^k_N \rangle$ and joint messages $\boldsymbol{m}^k = 
\langle m^k_1,..., m^k_N \rangle$, with $N=|I^k|$ denoting the number of agents in task $k$. The goal of multi-task MADRL with communication is to learn both the encoder $f$ and the policy $\pi$ that maximize the average expected return across all $K$ tasks:
$$
\mathcal{L}_{\mathcal{MT}} = \max_{\theta_{\pi}, \theta_{enc}} \frac{1}{K} \sum_{k=1}^K 
\mathbb{E}_{s^k \sim P^k,\, \boldsymbol{a}^k \sim \pi} 
\left[ \sum_{t=0}^T \gamma^{t} R^k(s^k_t, \boldsymbol{a}^k_t) \right]
$$
where $T$ is the length of episodes. Here, the policy $\pi$ is conditioned on local observations and messages. We follow the centralized training and decentralized execution (CTDE) paradigm to use a centralized value function as the critic to guide the training of decentralized policies. In practice, the centralized value function is learned from joint observations and messages to estimate expected return. For task $k$, we define a centralized observation-based value function $V(\boldsymbol{o}^k, \boldsymbol{m}^k; \phi)$ with parameters $\phi$ as:
$$
V(\boldsymbol{o}^k, \boldsymbol{m}^k; \phi) 
= \mathbb{E}_{s^k \sim P^k,\, \boldsymbol{a}^k \sim \pi} 
\Bigg[ \sum_{t = 0}^{T} \gamma^{t} 
R^k(s^k_t, \boldsymbol{a}^k_t) 
\,\Big|\, \boldsymbol{o}^k, \boldsymbol{m}^k, \phi \Bigg]
$$

\section{Methods}

\begin{figure}[t] 
    \centering
    \includegraphics[width=0.8\textwidth]{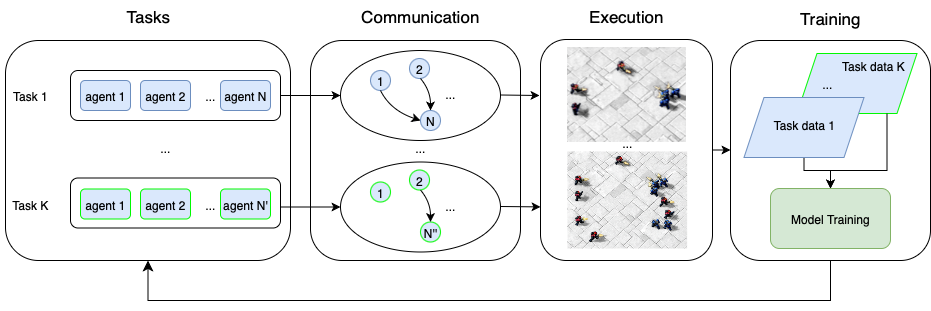}
    \caption{An overview of the MCS architecture. Agents communicate and act in each task, and data from multiple tasks are used to train a shared model across tasks.}
    \label{fig:overview}
\end{figure}

\begin{figure*}[t] 
    \centering
    \includegraphics[width=\textwidth]{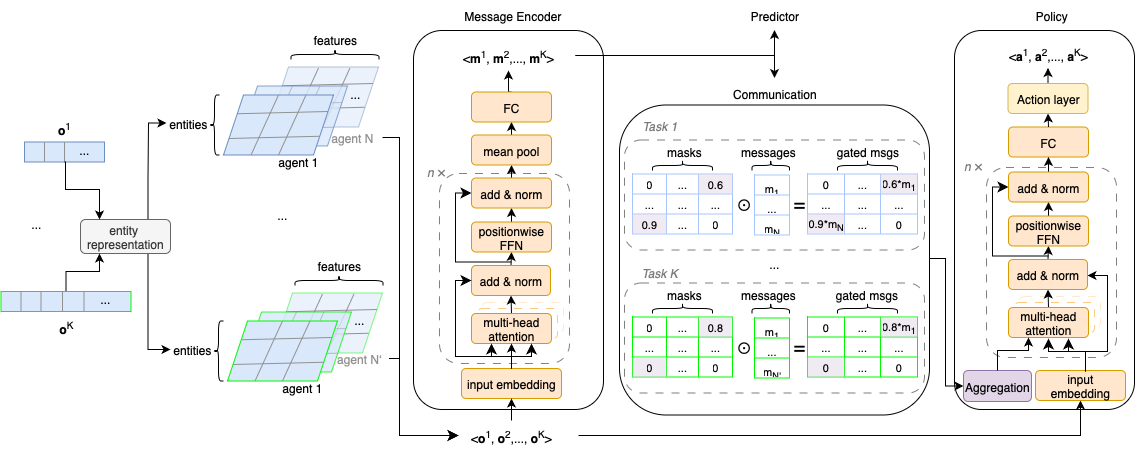}
    \caption{The network structure of MCS. Task-specific observations $\boldsymbol{o}^k$ are represented in an entity-based form and then encoded into messages $\boldsymbol{m}^k$. During communication, messages are pruned using masks $\boldsymbol{C}^k$, applied through column-wise multiplication. Then, messages are aggregated and integrated into the policy network.}
    \label{fig:modeldiagram}
\end{figure*}

In this section, we present Multi-task Communication Skills (MCS), a multi-task MADRL with communication method that learns and performs multiple tasks simultaneously. An overview of MCS architecture is shown in Figure \ref{fig:overview}. Each task consists of a set of agents that exchange messages through a learned communication graph. Based on received messages and local observations, agents select and execute actions in the environment. During training, experiences from all agents and tasks are collected jointly to enable centralized training. After training, agents deploy their learned policies in each individual task. MCS can handle communication with varying numbers of agents, observation spaces, and action spaces, thereby enabling generalization to multiple different environments. The entire framework, including communication, policy, and value networks, is trained end-to-end.

\subsection{Entity-based Communication in multi-task MADRL}
\label{sec:commMechanism}

We first introduce how messages are generated, communicated, and integrated into policies in MCS. The network structure used in MCS is schematically illustrated in Figure \ref{fig:modeldiagram}. We start with an entity-based representation across multiple tasks. Due to the varying dimensionality of the observation and action spaces, the network’s inputs and outputs shall be aligned when learning a shared communication protocol and policy across tasks. In entity-based representations, the observation $o^k_i$ for each agent $i$ in task $k$ is represented as $o^k_i \in \mathbb{R}^{|\hat{E}^k| \times D^e}$, where observable entities $\hat{E}^k$ may vary across tasks. However, the feature dimensionality $D^e$ for each entity remains constant across tasks (e.g., positions and velocity), which allows a shared input layer with dimensionality $D^e$ to be used for message generation. Next, given entity-based observations in task $k$, we employ a Transformer-based message encoder $f_{enc}$, shared by $N$ agents, to generate messages. Concretely, the message encoder $f_{enc}$ first embeds observations into an input embedding, followed by $n$ standard Transformer blocks with multi-head attention. The multi-head attention outputs a hidden representation with shape $\mathbb{R}^{|\hat{E}^k| \times D^h}$ for each agent $i$ in task $k$, where $D^h$ is the hidden dimensionality. To handle the varying sizes of entities across tasks and obtain low-dimensional messages, we apply mean pooling to aggregate features over entities, followed by a fully connected layer. This produces a message for agent $i$ in task $k$, denoted as $m^k_i = f_{enc}(o^k_i; \theta_{enc}) \in \mathbb{R}^{D^m}$, where $D^m$ is the dimensionality of messages that remains constant across tasks, and $\theta_{\text{enc}}$ represents the encoder parameters. As a result, the message encoder maps task- and agent-specific observations into a common message space $\mathbb{R}^{D^m}$, enabling the generalization of communication across agents and tasks. We denote the joint messages in task $k$ as $\boldsymbol{m}^k = \langle m^k_1, \dots, m^k_{N} \rangle = \langle f_{\text{enc}}(o^k_1; \theta_{\text{enc}}), \dots, f_{\text{enc}}(o^k_N; \theta_{\text{enc}}) \rangle$. This can be abbreviated as $\boldsymbol{m}^k = f_{\text{enc}}(\boldsymbol{o}^k; \theta_{\text{enc}})$.

We further prune unnecessary messages while leveraging an attention mechanism to measure the importance of communication between agents. By introducing a communication threshold, we construct a communication mask that prevents redundant messages and scales the remaining ones according to their importance. In particular, inspired by Zhang et al. \cite{Zhang2025TGCNet}, we employ an additive attention mechanism to produce scores $\alpha^{k}_{i,j}\in[0,1]$, which quantify the importance of agent $i$ communicating with agent $j$ in task $k$:
\begin{equation}
\alpha^k_{i,j} = \mathrm{Gumbel\text{-}Softmax}\!\left(
    v^\top \tanh \big( W_{q} m^{k}_i  + W_{k} m^{k}_j \big)
\right),
\label{eq:alpha}
\end{equation}
where $v^\top$, $W_{q}$, and $W_{k}$ are learnable parameters shared across agents and tasks. The communication mask $C^k_{i,j} \in [0,1]$ is defined based on the scores \footnote{We use soft differentiable samples via the Gumbel-Softmax to generate scores $\alpha^k_{ij}$.}:
\begin{equation}
C^k_{i,j} = 
\begin{cases}
\alpha^k_{i,j}, & \text{if } \alpha^k_{i,j} > \hat{\alpha} \\
0, & \text{otherwise}
\end{cases},
\quad i \neq j
\label{eq:commMask}
\end{equation}
where $\hat{\alpha}\in [0,1]$ is a predefined threshold. We denote gated messages received by agent $j$ from agent $i$ for task $k$ as $\tilde{m}^k_{i,j} = C^k_{i,j}\, m^k_i$. All messages received by agent $j$ in task $k$ are therefore denoted as $\tilde{\boldsymbol{m}}^{k}_j = \langle \tilde{m}^{k}_{1,j}, \dots, \tilde{m}^{k}_{N,j} \rangle = \boldsymbol{C}^k_j \odot \boldsymbol{m}^k$, where $\boldsymbol{C}^k_j = \langle C^k_{1,j}, \dots, C^k_{N,j} \rangle$ is the communication masks used by receiver agent $j$. In this way, the masks of redundant messages are set to 0 and therefore will not be considered in message integration.

Notably, the number of received messages can vary across tasks, which changes the input size of the policy network. To address this, we design a task-invariant architecture capable of aggregating a variable number of received messages while remaining efficient gradient backpropagation. We thus employ a GRU as an aggregation function $f_{agg}$, which can scale to a variable number of senders while capturing inter-agent dependencies. For each task $k$, the aggregation function produces an aggregated message $\bar{m}^{k}_{j}$ upon received messages $\tilde{\boldsymbol{m}}^{k}_j$ (i.e., $\bar{m}^{k}_{j}=f_{agg}(\tilde{\boldsymbol{m}}^{k}_j)$) as follows:
\begin{equation}
\begin{aligned}
h^{k}_{j,\ell} &= \mathrm{GRU}\!\big(\tilde m^{k}_{\ell,j},\, h^{k}_{j,\ell-1}\big),
\qquad \ell=1,\dots,N\\
\bar{m}^{k}_{j} &= \frac{1}{N}\sum_{\ell}{h^{k}_{j,\ell}} \in R^{D_h},
\end{aligned}
\label{eq:aggregate}
\end{equation}
where $h^{k}_{j,\ell}$ is the hidden state of receiver agent $j$  in task $k$ when processing the $\ell$-th received message.

The aggregated message $\bar{m}^k_j$ is then used to help receiver agents focus on important and relevant features in task-specific observations when deciding an action. Specifically, we use $\bar{m}^k_j$ as the query embedding in a standard multi-head attention module within the Transformer-based policy network. Let $o_j^k\in\mathbb{R}^{|\hat{E}^k|\times D^e}$ denote the observation matrix for agent $j$ in task $k$. For $\tilde{H}$ heads with $d_{\text{head}}=D^h/\tilde{H}$, each head $\tilde{h}\in\{1,\dots,\tilde{H}\}$ computes:
\begin{equation}
\begin{aligned}
& \mathcal{Q}_{j}^{(\tilde{h})} = (\bar m^{k}_j)^\top W_{\mathcal{Q}}^{(\tilde{h})} \in \mathbb{R}^{1\times d_{\text{head}}},\\
& \mathcal{K}_{j}^{(\tilde{h})} = o_j^k W_{\mathcal{K}}^{(\tilde{h})} \in \mathbb{R}^{|\hat{E}^k|\times d_{\text{head}}} \\
& \mathcal{V}_{j}^{(\tilde{h})} = o_j^k W_{\mathcal{V}}^{(\tilde{h})} \in \mathbb{R}^{|\hat{E}^k|\times d_{\text{head}}},\\
& \mu_{j}^{(\tilde{h})} = \operatorname{softmax}\!\left(\frac{\mathcal{Q}_{j}^{(\tilde{h})} {\mathcal{K}_{j}^{(\tilde{h})}}^\top}{\sqrt{d_{\text{head}}}}\right) \in \mathbb{R}^{1\times |\hat{E}^k|},\\
& z^k_j = \big[\,\!\big\|_{\tilde{h}=1}^{\tilde{H}} \mu_{j}^{(\tilde{h})} \mathcal{V}_{j}^{(\tilde{h})} \big]\, W_{z} \;\in\; \mathbb{R}^{1\times D^h}
\end{aligned}
\label{eq:msgQKV}
\end{equation}
where $W_{\mathcal{Q}}$, $W_{\mathcal{K}}$, $W_{\mathcal{V}}$, and $W_{z}$ are learnable weight matrices shared across agents and tasks. Notably, the query embedding $\mathcal{Q}{j}^{(\tilde{h})}$ and the key embedding $\mathcal{K}{j}^{(\tilde{h})}$ are used to produce weighting scores $\mu_{j}^{(\tilde{h})}$ over entities, which indicate the relevance of the received message to each observable entity of receiver $j$. These weighting scores are then used to combine the entity-based representations derived from observations. The outputs from all heads are concatenated to form $z^k_j$, which enables receiver $j$ to focus on the most relevant entity-based observations during communication. Then, $z^k_j$ is used in the action layer to decide the receiver's actions. Note that the final action layer to generate task-specific actions follows Hu et al. \cite{Hu2021UPDet} and Tian et al. \cite{Tian2024DT2GS}. In MCS, messages are used not only in the policy network but also as additional inputs to the critic networks. Then, each sender agent’s messages are updated through gradient backpropagation from both the receiver agents’ policy networks and the critic networks, thereby guiding the generation of messages that are most beneficial for communication. Within CTDE paradigm, we optimize the following objective across all agents and tasks:

\begin{equation}
\begin{aligned}
&\mathcal{L}(\theta_{enc}, \theta_{\pi}, \phi)= \\
&\frac{1}{K}\sum_k^K\frac{1}{N}\sum_i^{N}\mathbb{E}_{\boldsymbol{o}^k,a^k_i}[\log \pi(a^k_i|o^k_i,\bar{m}^k_i; \theta_{\pi})V(\boldsymbol{o}^k, \bar{\boldsymbol{m}}^k; \phi)]
\end{aligned}
\label{eq:commObj}
\end{equation}
where $\bar{\boldsymbol{m}}^k=\langle \bar{m}^k_1,\dots,\bar{m}^k_{N} \rangle$ denotes the joint aggregated messages of agents, and $\bar{m}^k_i = f_{agg}\!\left(\boldsymbol{C}^k_i\odot f_{enc}(\boldsymbol{o}^k;\theta_{enc})\right)$. During centralized training, the encoder parameters $\theta_{enc}$, policy parameters $\theta_{\pi}$, and critic parameters $\phi$ are shared across agents and tasks.

\subsection{Prediction Network for Coordination}
\label{sec:predictorNet}

\begin{figure}[t] 
    \centering
    \includegraphics[width=0.7\textwidth]{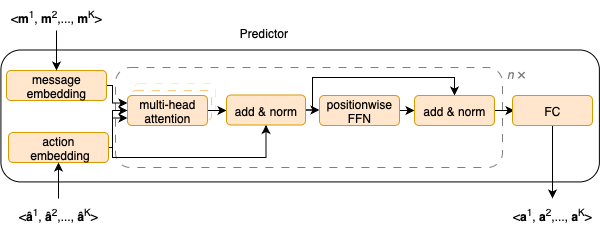}
    \caption{The network structure diagram of the predictor.}
    \label{fig:predictor}
\end{figure}

With our proposed message encoder, communication is used to share encoded task-specific observations among agents, thereby broadening each agent’s perspective of the overall environment. Despite this, sharing encoded observations does not explicitly account for coordination, as agents may still need to align their action selections to achieve coordinated behaviors. To address this, we further introduce a prediction network $q_{pred}$ that correlates the generated messages $\boldsymbol{m}^k$ with the sender agents’ actions $\boldsymbol{a}^k$, thereby enhancing coordinated action selection among agents. Concretely, we employ an additional Transformer decoder as the prediction network (Figure \ref{fig:predictor}), which is used only during training and discarded at execution. Within multi-head attention, the prediction network takes received messages as query embeddings. The key and value embeddings are derived from initialized actions denoted as $\tilde{\boldsymbol{a}}$ (e.g., null actions). To handle varying action dimensionalities across tasks, we apply zero-padding to actions. The predicted action distribution is then used for maximizing the mutual information $I(A^k;M^k)$ between actions $\boldsymbol{a}^k \in A^k$ and messages $\boldsymbol{m}^k \in M^k$ for each task $k$. However, computing the mutual information $I(A^k;M^k)$ is intractable. Therefore, we maximize its variational lower bound (known as Barber-Agakov lower bound) \cite{Poole2019Variation}. By replacing the conditional distribution of actions given messages with a variational distribution $q_{pred}(\boldsymbol{a}^k \mid \boldsymbol{m}^k)$, we have:

\begin{equation}
\begin{aligned}
&I(A^k;M^k) = \mathbb{E}_{p(\boldsymbol{a}^k,\boldsymbol{m}^k)}\!\left[\log \frac{p(\boldsymbol{a}^k\mid \boldsymbol{m}^k)}{p(\boldsymbol{a}^k)}\right] \\
&= \mathbb{E}_{p(\boldsymbol{a}^k,\boldsymbol{m}^k)}\!\left[\log \frac{q_{pred}(\boldsymbol{a}^k\mid \boldsymbol{m}^k)\,p(\boldsymbol{a}^k\mid \boldsymbol{m}^k)}{p(\boldsymbol{a}^k)\,q_{pred}(\boldsymbol{a}^k\mid \boldsymbol{m}^k)}\right] \\
&= \mathbb{E}_{p(\boldsymbol{a}^k,\boldsymbol{m}^k)}\!\left[\log \frac{q_{pred}(\boldsymbol{a}^k\mid \boldsymbol{m}^k)}{p(\boldsymbol{a}^k)}\right]
  \;+ \\
&\; \mathbb{E}_{p(\boldsymbol{m}^k)}\!\Big[\mathrm{KL}\!\big(p(\boldsymbol{a}^k\mid \boldsymbol{m}^k)\,\|\,q_{pred}(\boldsymbol{a}^k\mid \boldsymbol{m}^k)\big)\Big] \\
&\ge \mathbb{E}_{p(\boldsymbol{a}^k,\boldsymbol{m}^k)}\!\big[\log q_{pred}(\boldsymbol{a}^k\mid \boldsymbol{m}^k)\big] \;+\; H(\boldsymbol{a}^k)
\end{aligned}
\label{eq:deriveObj}
\end{equation}
where $H(\boldsymbol{a}^k)\equiv -\,\mathbb{E}_{p(\boldsymbol{a}^k)}[\log p(\boldsymbol{a}^k)]$ is the entropy of action distribution. The last inequality is due to the non-negativity of the KL divergence. Since the entropy term is non-negative, maximizing the first term is equal to maximizing the mutual information. In practice, we estimate $\mathbb{E}_{p(\boldsymbol{a}^k,\boldsymbol{m}^k)}\!\big[\log q_{pred}(\boldsymbol{a}^k\mid \boldsymbol{m}^k)\big]$ using sampled batch data. Given samples $\{(\boldsymbol{a}^k_b, \boldsymbol{m}^k_b, \boldsymbol{o}^k_b)\}_{k \in \{1,...,K\}, b \in \{1,...,B\}}$ for $K$ tasks and $B$ batch-size, we maximize the following log-likelihood:
\begin{equation}
\mathcal{L}_{pred}(\theta_{enc})
= \,\frac{1}{K} \frac{1}{B} \sum_{k=1}^{K} \sum_{b=1}^{B}
\log q_{pred}\!\Big(\boldsymbol{a}^k_b \,\Big|\, f_{\text{enc}}\!\big(\boldsymbol{o}^k_b;\theta_{enc}\big)\Big)\qquad
\label{eq:predObj}
\end{equation}
The loss $\mathcal{L}_{pred}$ backpropagates gradients from the predictor to the message encoder, encouraging it to generate messages that reflect the sender agents’ behaviors and align agents' action selections.

\subsection{The Overall Learning Objective}

The overall learning objective $\mathcal{L}_{\mathcal{MT}}$ of multi-task MADRL with communication is defined as:
\begin{equation}    
\mathcal{L}_{\mathcal{MT}} = \mathcal{L}(\theta_{enc}, \theta_{\pi}, \phi) + \beta \,\mathcal{L}_{pred}(\theta_{enc}),
\label{eq:totalLoss}
\end{equation}
where $\beta$ is a coefficient that balances the influence of the prediction network. $\mathcal{L}_{\mathcal{MT}}$ ensures that messages are both informative and beneficial for action selections, and that the policies and critics are learned based on communication. In particular, the messages are optimized not only through gradients backpropagated from the receiver agents’ policy and critic networks, but also through the auxiliary prediction network.

We further introduce Algorithm \ref{alg:algorithmOne} to illustrate how MADRL agents communicate and update their messages, policies, and critics in the multi-task setting. At each time step $t$ of an episode, agents in task $k$ observe $\boldsymbol{o}^k_t=\langle o^k_{1,t},\dots,o^k_{N,t} \rangle$, which are used to generate messages $\boldsymbol{m}^k_t$ (lines 5-6). Based on the communication graph defined in Equations \ref{eq:alpha} and \ref{eq:commMask} and the aggregation function in Equation \ref{eq:aggregate}, messages are communicated and aggregated into $\bar{\boldsymbol{m}}^k_t$ at time step $t$ (lines 7-9), which are then used to produce actions $\boldsymbol{a}^k_t$ (line 10). The actions are executed in the environment, and rewards $r^k_t$ are collected (line 10). The resulting transitions of observations, actions, and rewards are stored in a task-specific replay buffer (line 11). During training, mini-batches are sampled from these buffers (line 14). Lines 15–17 are then used to compute the losses for the policy, predictor, and critic in Equation \ref{eq:totalLoss}. Concretely, the sampled observations are first used to generate messages, which are subsequently passed through the prediction network to produce an action distribution. Then, gradients are backpropagated through the critics, predictors, actors, aggregation function, and message encoders in an end-to-end manner (line 20). 


\begin{algorithm}[t]
\caption{Multi-task Communication Skills}\label{alg:algoGAAC}
\begin{algorithmic}[1]
\State \textbf{Input}: Batch size $B$, number of tasks $K$, episodes $L$, steps $T$.
\State \textbf{Initialize}: Encoder parameters $\theta_{enc}$, policy parameters $\theta_{\pi}$, and critic parameters $\phi$. Replay buffer $\mathcal{B}^k$ for each task $k$.
\For{$l= 0,1,...,L - 1$}
\For{$t= 0,1,...,T - 1$}
\State Collect observations $(\boldsymbol{o}^1_t,...,\boldsymbol{o}^K_t)$ for $K$ tasks
\State Generate messages $(\boldsymbol{m}^1_t,...,\boldsymbol{m}^K_t)$ with message encoder
\For{$k= 1,...,K$}
\State Communicate and aggregate messages into $\bar{\boldsymbol{m}}^k_t$ based on Equations \ref{eq:alpha}, \ref{eq:commMask}, \ref{eq:aggregate}
\EndFor
\State Decide actions $(\boldsymbol{a}^1_t,...,\boldsymbol{a}^K_t)$ based on $(\bar{\boldsymbol{m}}^1_t,...,\bar{\boldsymbol{m}}^K_t)$ in Equation \ref{eq:msgQKV} and collect the rewards $(r^1_t,...,r^K_t)$
\State Insert $(\boldsymbol{o}^k_t,\boldsymbol{a}^k_t,r^k_t)$ into buffer $\mathcal{B}^k$
\EndFor
\For{$k= 1,...,K$}
\State Sample a random minibatch of $B$ steps from $\mathcal{B}^k$
\State Generate messages $\boldsymbol{m}^k=f_{enc}(\boldsymbol{o}^k)$
\State Generate predicted action distribution $q_{pred}(\boldsymbol{a}^k|\boldsymbol{m}^k)$
\State Generate critics $V(o^k_1, \bar{m}^k_1),...,V(o^k_{N}, \bar{m}^k_{N})$
\EndFor
\State Calculate loss $-\mathcal{L}_{\mathcal{MT}}$ based on Equation \ref{eq:totalLoss}
\State Update critics, actors, and messages with gradient descent
\EndFor
\end{algorithmic}
\label{alg:algorithmOne}
\end{algorithm}


\section{Experiments}

We evaluate our proposed method in three challenging multi-agent environments: AliceBob \cite{Yang2023HMASD}, SMAC \cite{Samvelyan2019SMAC}, and Google Research Football \cite{Kurach2020Football} \footnote{The source code will be released publicly on Github upon acceptance.}. These environments were originally designed for single-task MADRL and consist of multiple cooperative tasks with challenges in coordinating agents' behaviors. Following Tian et al. \cite{Tian2024DT2GS}, we construct multi-task settings for each environment. We then compare the following methods across these multiple tasks:

\begin{itemize}[leftmargin=*]
    \item \textbf{Multi-task MADRL methods with communication}. Our proposed method, MCS, is the first in this branch, which is a Transformer-based approach that allows communication among agents for better coordination in multiple tasks. In MCS, $K$ tasks are learned and evaluated simultaneously. 
    
    \item \textbf{Multi-task MADRL methods without communication}. DT2GS \cite{Tian2024DT2GS} and RIT \cite{Li2025RIT} are two SOTA methods in this branch. DT2GS is a policy-based method which uses Transformers as an encoding of entity-based observations and decode actions for multiple tasks. RIT is a value-based method which uses padding techniques for unobservable entities for different tasks. We compare MCS with DT2GS and RIT to evaluate the benefit of communication under multi-task settings. Similar to MCS, $K$ tasks are learned and evaluated simultaneously. 

    \item \textbf{Single-task MADRL methods with communication}. TGCNet \cite{Zhang2025TGCNet} is the most recent method that employs a multi-key gated attention mechanism. In contrast to MCS, TGCNet does not consider the correlation between messages and actions. MAIC \cite{Yuan2022MAIC} is another communication approach that promotes coordination by sharing messages correlated with other agents’ intended actions. Notably, neither TGCNet nor MAIC considers the multi-task setting. Instead, $K$ tasks are learned and evaluated separately.
    
    \item \textbf{Single-task MADRL methods without communication}. MAT \cite{Wen2022MAT} is the SOTA method on many SMAC tasks, which employs a Transformer to encode observations and generate action sequences for agents. HMASD \cite{Yang2023HMASD} represents the SOTA in leveraging latent skills for agent coordination. We compare MCS with MAT to evaluate the impact of our proposed Transformer-based communication mechanism, and with HMASDA to assess the effectiveness of the prediction network in promoting coordination. Also, $K$ tasks are learned and evaluated separately.

\end{itemize}

The comparison examines two key aspects: (i) whether training is conducted in a multi-task or single-task setting, and (ii) whether communication is enabled among agents. This results in two solid perspectives for comparison: Multi-task with vs. without communication, and Multi-task vs. Single-task. By following the literature \cite{Tian2024DT2GS}, we evaluate the averaged performance across $K$ tasks, defined as $Avg=\frac{1}{K} \sum_k^K WinRate(k)$, where $WinRate(k)$ denotes the win rate on task $k$. The win rate $WinRate(k)$ is estimated by running several evaluation episodes for each task at fixed intervals during training, then averaging across episodes and tasks. We further show task-specific performance $WinRate(k)$ for each method, which demonstrates whether multi-task methods can outperform sing-task methods in each particular task. The hyper-parameters of each methods are either from published papers or fine-tuned for fair comparison. For MCS, we further fine-tune the threshold $\hat{\alpha}$ (Equation \ref{eq:commMask}) and coefficient $\beta$ (Equation \ref{eq:totalLoss}), and we also analyze the sensitivity of these parameters on the learning performance. We provide a list of critical hyperparameters and the details of the entity-based representation for each environment in Appendix.

\subsection{Evaluation Results}
\label{sec:evalRes}

\begin{figure*}[t]
  \centering
  \begin{subfigure}{0.19\textwidth}
    \centering
    \includegraphics[width=\linewidth]{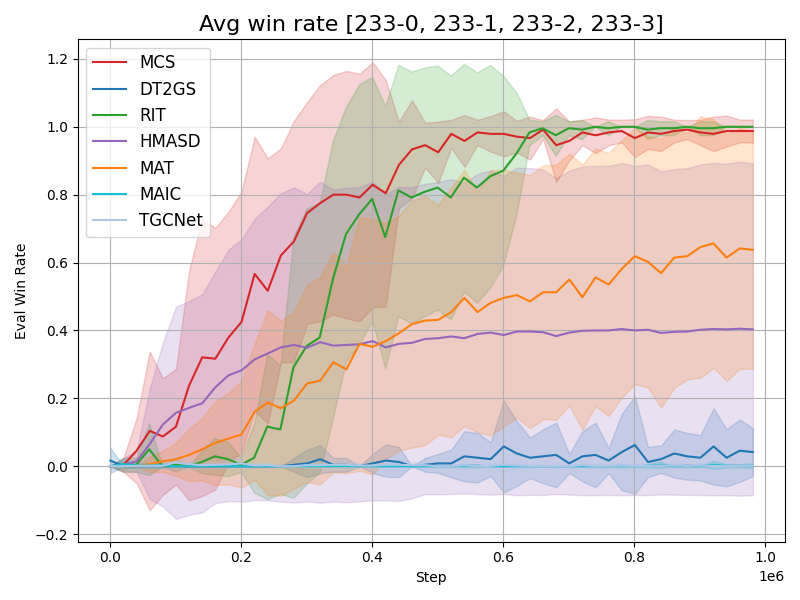}
    \caption{AliceBob-233}
    \label{fig:main-alice-233}
  \end{subfigure}
  \begin{subfigure}{0.19\textwidth}
    \centering
    \includegraphics[width=\linewidth]{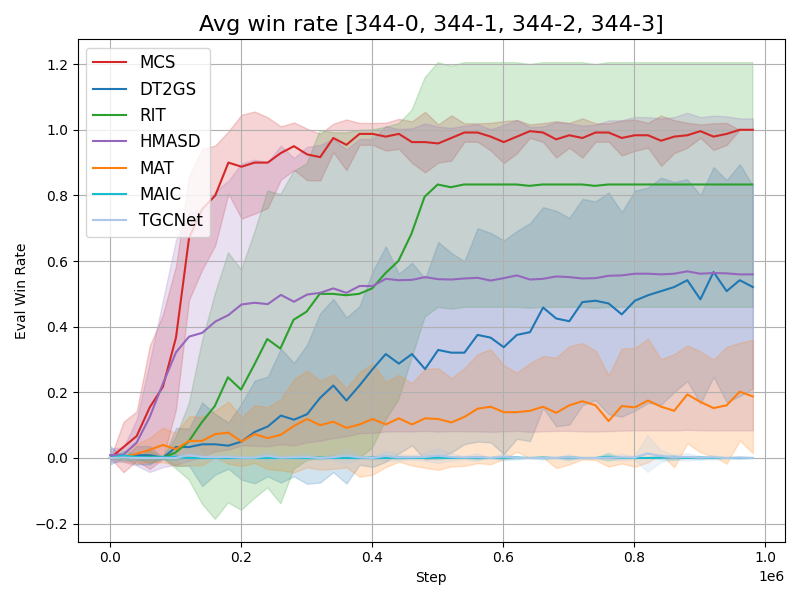}
    \caption{AliceBob-344}
    \label{fig:main-alice-344}
  \end{subfigure}
  \begin{subfigure}{0.19\textwidth}
    \centering
    \includegraphics[width=\linewidth]{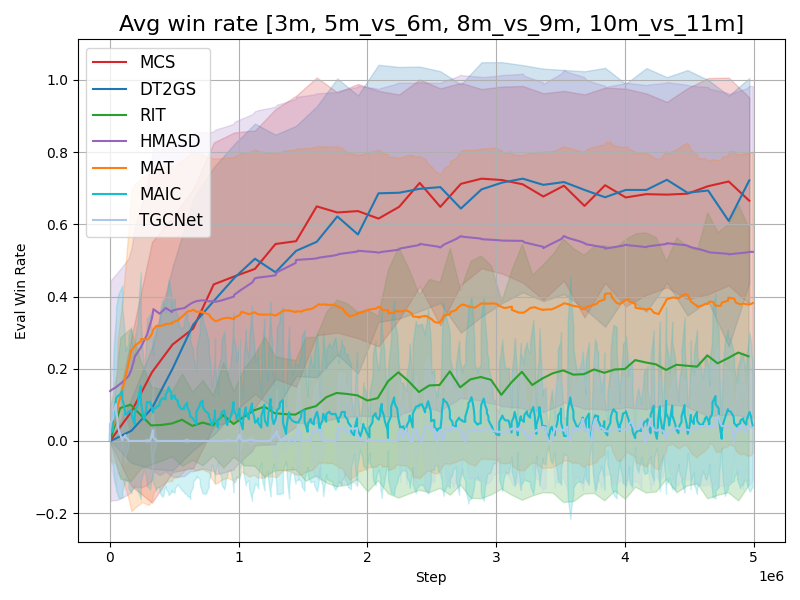}
    \caption{Marine}
    \label{fig:main-marine}
  \end{subfigure}
  \begin{subfigure}{0.19\textwidth}
    \centering
    \includegraphics[width=\linewidth]{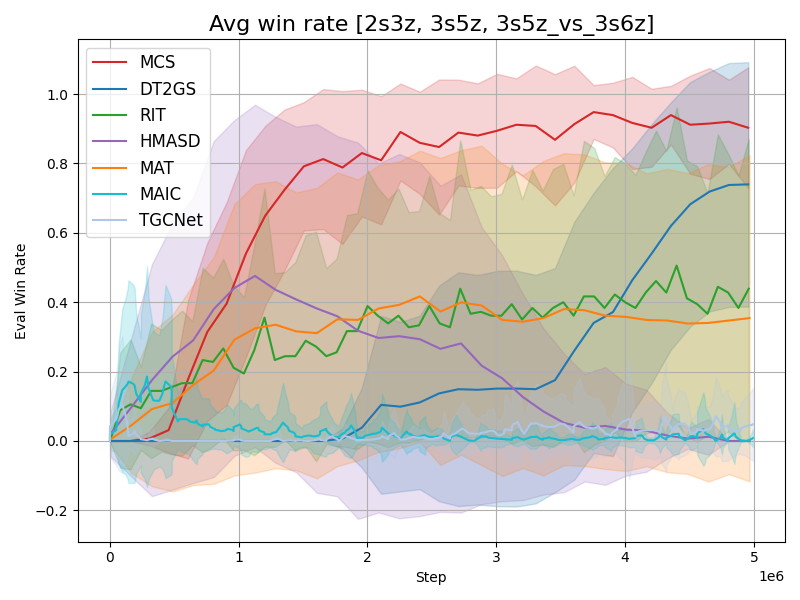}
    \caption{Stalker–Zealot}
    \label{fig:main-stalker-zealot}
  \end{subfigure}
  \begin{subfigure}{0.19\textwidth}
    \centering
    \includegraphics[width=\linewidth]{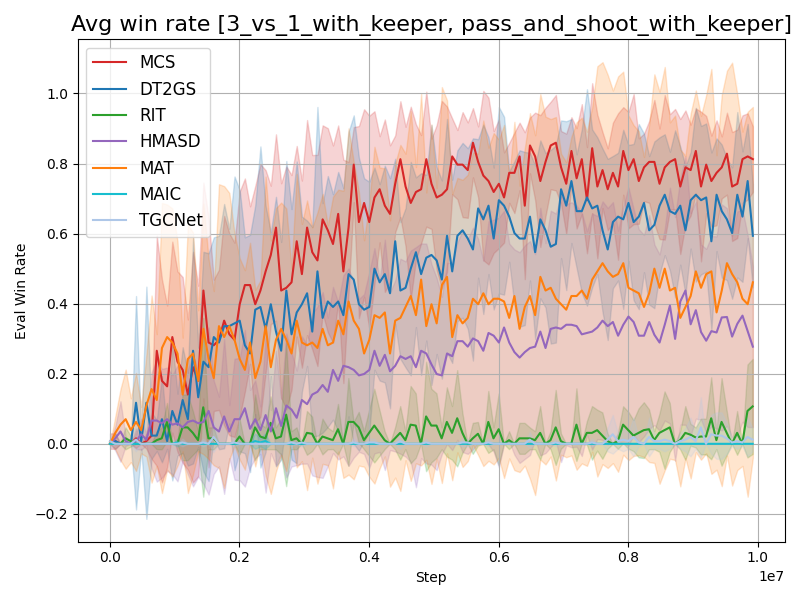}
    \caption{Football}
    \label{fig:main-football}
  \end{subfigure}
  \caption{Averaged win-rate across multiple tasks on AliceBob (a–b), SMAC (c–d), and Football (e).}
  \label{fig:winrate_average}
\end{figure*}

\begin{figure*}[t]
  \centering

  \begin{subfigure}{0.18\linewidth}
    \includegraphics[width=\linewidth]{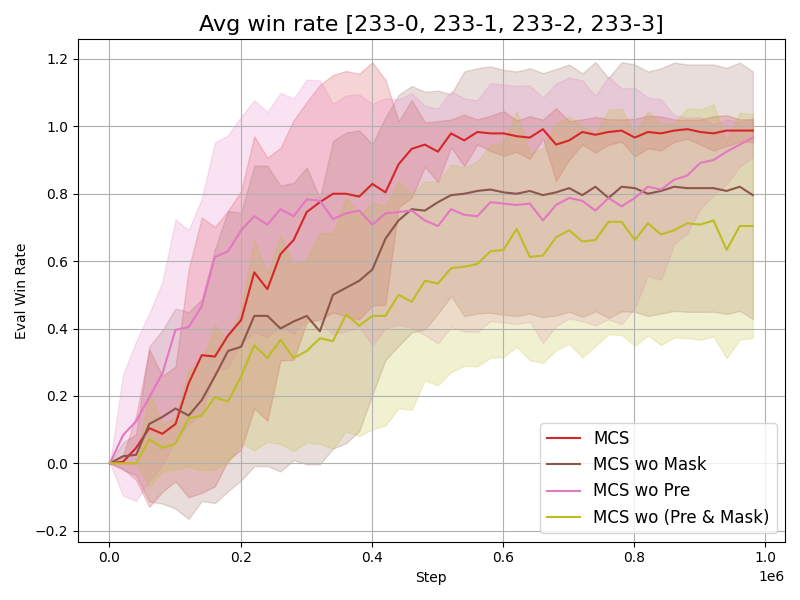}
    \caption{AliceBob-233}
    \label{fig:abl-alice-233}
  \end{subfigure}\hfill
  \begin{subfigure}{0.18\linewidth}
    \includegraphics[width=\linewidth]{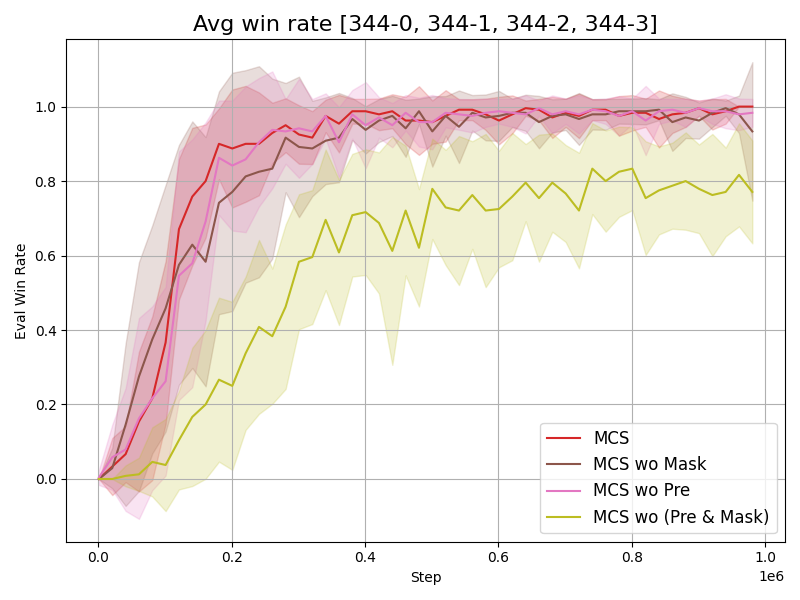}
    \caption{AliceBob-344}
    \label{fig:abl-alice-344}
  \end{subfigure}\hfill
  \begin{subfigure}{0.18\linewidth}
    \includegraphics[width=\linewidth]{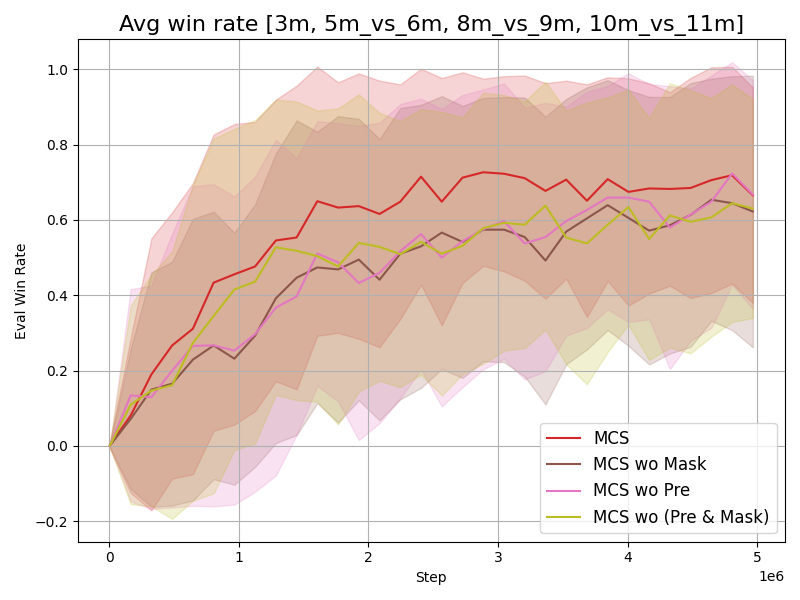}
    \caption{Marine}
    \label{fig:abl-smac-Marine}
  \end{subfigure}\hfill
  \begin{subfigure}{0.18\linewidth}
    \includegraphics[width=\linewidth]{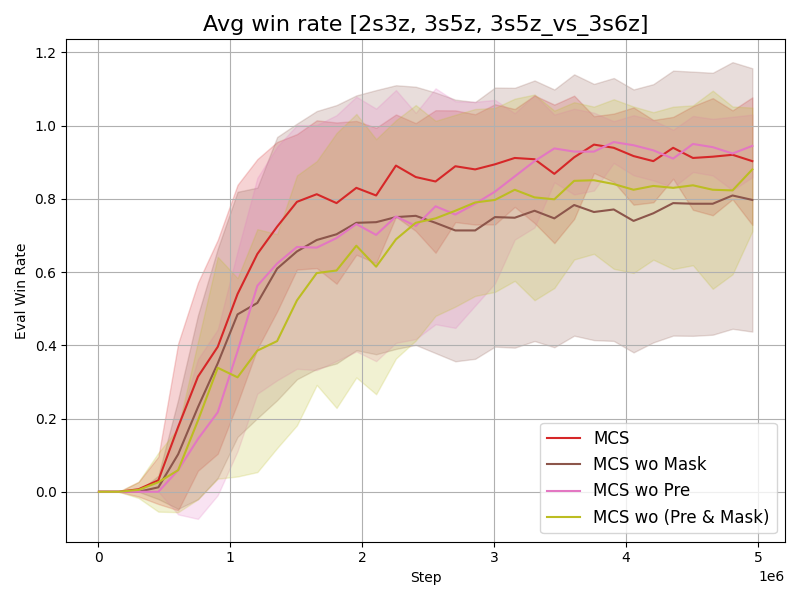}
    \caption{Stalker-Zealot}
    \label{fig:abl-smac-sz}
  \end{subfigure}\hfill
  \begin{subfigure}{0.18\linewidth}
    \includegraphics[width=\linewidth]{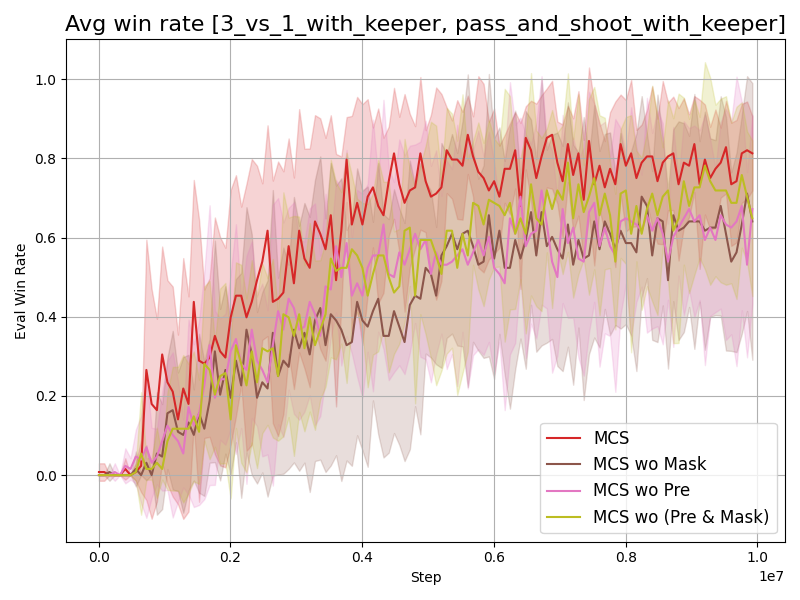}
    \caption{Football}
    \label{fig:abl-football}
  \end{subfigure}

  \caption{Ablation studies of MCS on AliceBob (a–b), SMAC (c–d), and Football (e).}
  \label{fig:ablCompare}
\end{figure*}

\paragraph{\textnormal{\textbf{AliceBob Series}}} The AliceBob environment, originally designed by Yang et al. \cite{Yang2023HMASD}, is created to explicitly demonstrate how agents coordinate to collect diamonds in a grid world, where a diamond can only be collected if another agent is simultaneously standing on the button of the same color. We extend this environment to a multi-task setting where each task consists of either diamonds or food, and either buttons or keys. For example, task $0$ may involve 3 pairs of entities $\{red\,diamond,\, red\,button\}, \{blue\,diamond,$ $blue\,button\}, \{pink\,diamond,\, pink\,button\}$ with 2 agents $\{Alice, Bob\}$. We denote this configuration as $233$-$0$, indicating 2 agents, 3 diamonds/foods, and 3 buttons/keys with index $0$. By varying entity types and associated colors (therefore different IDs), we obtain the AliceBob-$233$ series: $\{233$-$0, 233$–$1$, $233$-$2, 233$–$4\}$. To examine scalability, we further construct the AliceBob-$344$ series $\{344$–$0, 344$–$1, 344$–$2, 344$-$4\}$ by introducing an additional agent (Charlie) as well as an additional color and entity (e.g., blue flower). To increase task complexity, diamonds and trees are randomly placed at the top of the environment, while keys and buttons are randomly placed at the bottom at the beginning of each episode. Agents are also spawned at random positions, requiring them to coordinate effectively to reach the correct and dynamically changing targets. Each agent perceives only the relative distances between its own position and other entities, and its surrounding grids. The reward function is defined as follows: agents receive a reward of 1 whenever a valid pair (diamond–button or food–key) is completed, a reward of 5 when all pairs are completed (regarded as a win), a penalty of 0.5 for agent collisions, and a penalty of 0.1 per step. 

We show the averaged performance across tasks in AliceBob-$233$ and AliceBob-$344$ series in Figures \ref{fig:main-alice-233}-\ref{fig:main-alice-344}, with results averaged over 6 random seeds. MCS shows faster convergence than other methods in AliceBob-$233$ series and significantly better performance than other methods in AliceBob-$344$ series. With communication, the Charlie agent in AliceBob-$344$ series can be used for better reaching the goals in MCS. In contrast, RIT, which applies a unified domain mask across tasks, achieves comparable performance in this environment, suggesting that it can effectively generalize knowledge between tasks. DT2GS, however, does not consider inter-agent coordination and therefore fails in the tasks. Other methods such as TGCNet and MAIC rely on training with single batch, which prevents them from efficiently exploiting successful episodes (where agents must collect all diamonds or food for a win). We also report per-task performance for both multi-task and single-task methods in the Appendix, where MCS consistently achieves much higher win rate across all tasks in AliceBob 344 series.

\paragraph{\textnormal{\textbf{SMAC Series}}.} The StarCraft Multi-Agent Challenge (SMAC) is a real-time strategy game serving as a benchmark in the MADRL community \cite{Samvelyan2019SMAC}. In SMAC, agents controlled by the learning algorithm must defeat all enemies, requiring effective cooperation strategies and fine-grained micro-control of movement and attack. Following Tian et al.~\cite{Tian2024DT2GS}, we construct two series of maps: the Marine series $\{3m, 5m\_vs\_6m, 8m\_vs\_9m, 10m\_vs\_11m\}$ and the Stalker–Zealot series $\{2s3z, 3s5z, 3s5z\_vs\_3s6z\}$. In these settings, the number of agents varies across tasks, testing the scalability of multi-task methods. Moreover, the observation and action spaces differ across tasks, adding further complexity to learn a unified model. 

Figures \ref{fig:main-marine}-\ref{fig:main-stalker-zealot} reports the average performance across the Marine and Stalker–Zealot series, with results averaged over 6 random seeds. On the Marine series, MCS achieves similar performance as DT2GS, and both MCS and DT2GS outperform other baselines, including the strong single-task method HMASD. On the Stalker–Zealot series, MCS significantly outperforms all the other methods. Notably, HMASD suffers a severe performance drop around 1M time steps, indicating convergence to a poor local optimum. RIT converges very quickly but with much lower win rates. DT2GS improves only after about 3M steps, which is much slower than MCS in this series. We further report per-task performance for both multi-task and single-task methods in the Appendix, where MCS consistently achieves the highest win rates across all tasks in the Stalker–Zealot series.

\paragraph{\textnormal{\textbf{Football Series}}.} 

Google Research Football (GRF) is a physics-based 3D soccer simulator for reinforcement learning \cite{Kurach2020Football}, providing a challenging multi-agent benchmark environment with high stochasticity and sparse rewards. In GRF, opponents are controlled by expert agents, which significantly increase the complexity of the state–action space. To highlight the difficulty of the tasks, we train and evaluate using only the scoring reward: agents receive a reward of +1 when scoring a goal, which is sparse. Episodes terminate when a goal is scored, the ball goes out of bounds, or possession changes. To construct entity-based representations, features from the left and right teams are grouped into four macro entities: left-team locations, left-team directions, right-team locations, and right-team directions. Ball position and direction are appended to the appropriate entity depending on possession, and a one-hot encoding of the active player is also included. The final entity representation remains comparable in size to the original observations, which does not alter task difficulty. We construct a multi-task Football series using two maps: $\{3\_vs\_1\_with\_keeper, \; pass\_and\_shoot\_with\_keeper\}$, which involve different numbers of RL agents attempting to score from the edge of the field. 

Figure \ref{fig:main-football} shows the averaged win rate across tasks in the Football series, with results further averaged over 4 random seeds. MCS achieves significantly a higher averaged win rate than all baseline methods. RIT fails to perform effectively in this high-dimensional, stochastic domain. The results of per-task performance are provided in the Appendix, where MCS obtains the highest win rate in $3\_vs\_1\_with\_keeper$ and achieves similar performance as MAT and DT2GS in $pass\_and\_shoot\_with\_keeper$.

\subsection{Ablation Studies}
\label{sec:ablStu}

We conduct ablation experiments on MCS to assess the contributions of the communication mask (Equation \ref{eq:commMask}) and the predictor (Equation \ref{eq:predObj}), investigating whether pruning unnecessary messages and encouraging the correlation between messages and actions improves learning performance. As can be seen from Figure \ref{fig:ablCompare}, removing either communication mask (MCS wo Mask) defined in Equation \ref{eq:commMask} or the prediction network (MCS wo Pre) defined in Section \ref{sec:predictorNet} can lead to slower convergence or performance degradation. In all multi-task series, removing both communication mask and prediction network  (MCS wo (Pre \& Mask)) can lead to significantly lower win rate compared to MCS.

\subsection{Hyperparameters Analysis}

\begin{figure*}[t]
  \centering

  \begin{subfigure}{0.15\linewidth}
    \includegraphics[width=\linewidth]{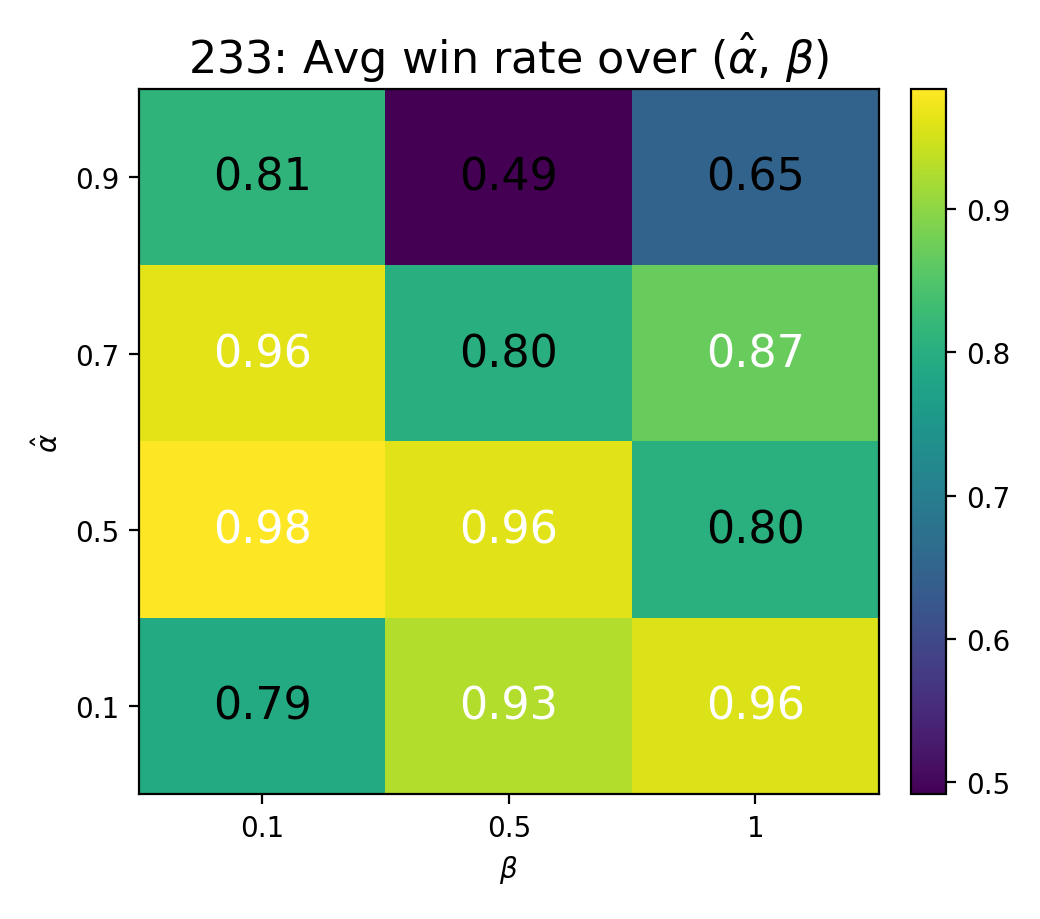}
    \caption{AliceBob-233}
    \label{fig:param-alice-233}
  \end{subfigure}\hfill
  \begin{subfigure}{0.15\linewidth}
    \includegraphics[width=\linewidth]{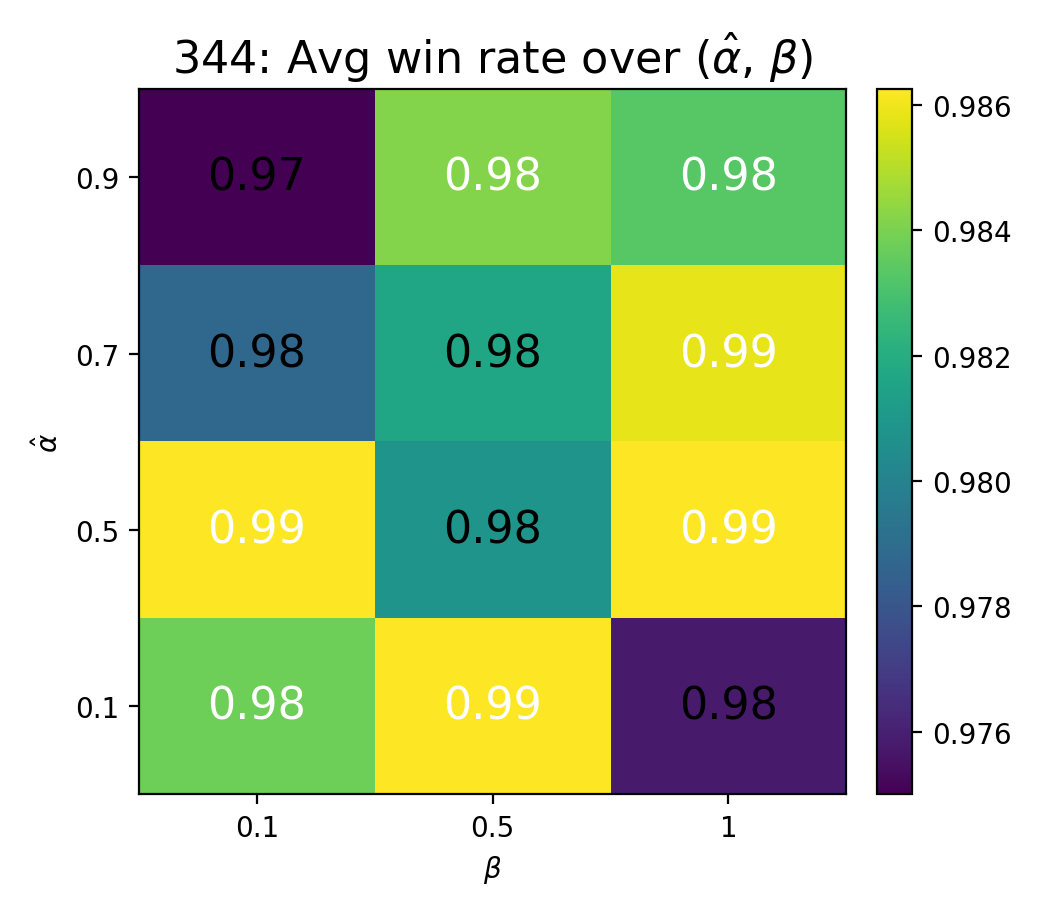}
    \caption{AliceBob-344}
    \label{fig:param-alice-344}
  \end{subfigure}\hfill
  \begin{subfigure}{0.15\linewidth}
    \includegraphics[width=\linewidth]{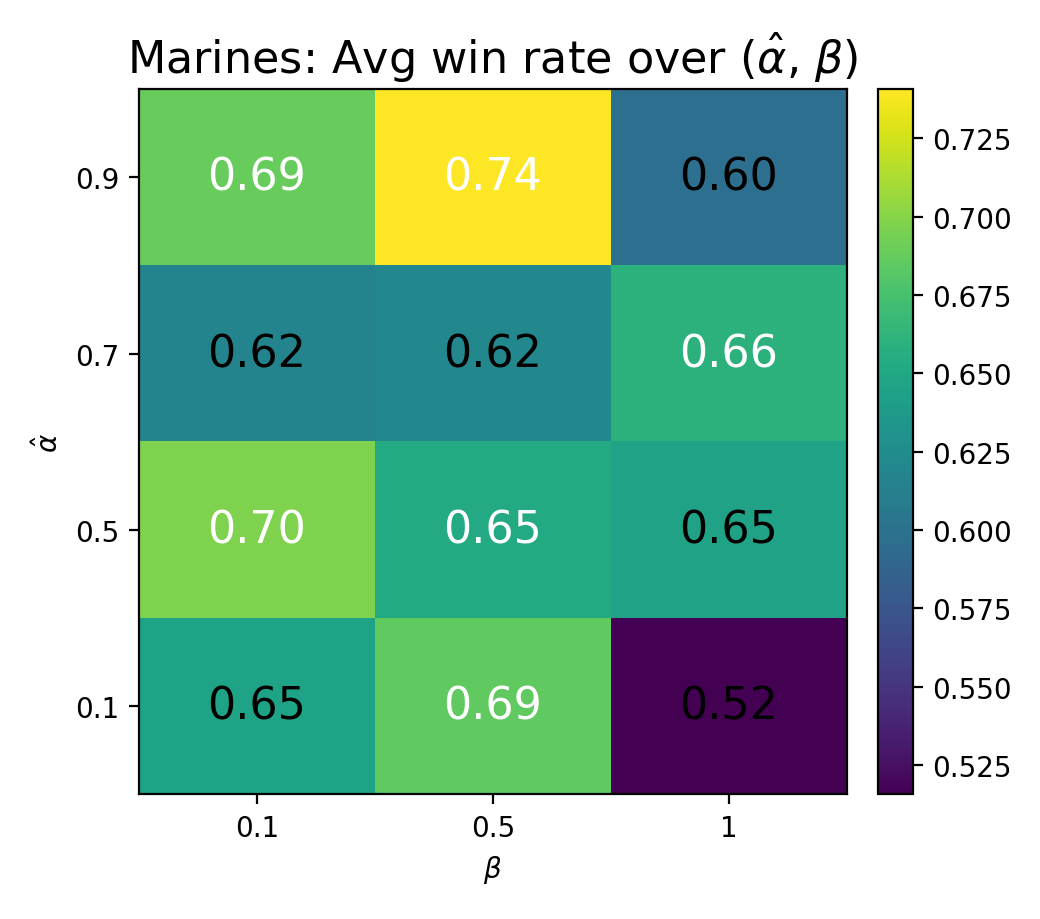}
    \caption{Marine}
    \label{fig:param-smac-Marine}
  \end{subfigure}\hfill
  \begin{subfigure}{0.15\linewidth}
    \includegraphics[width=\linewidth]{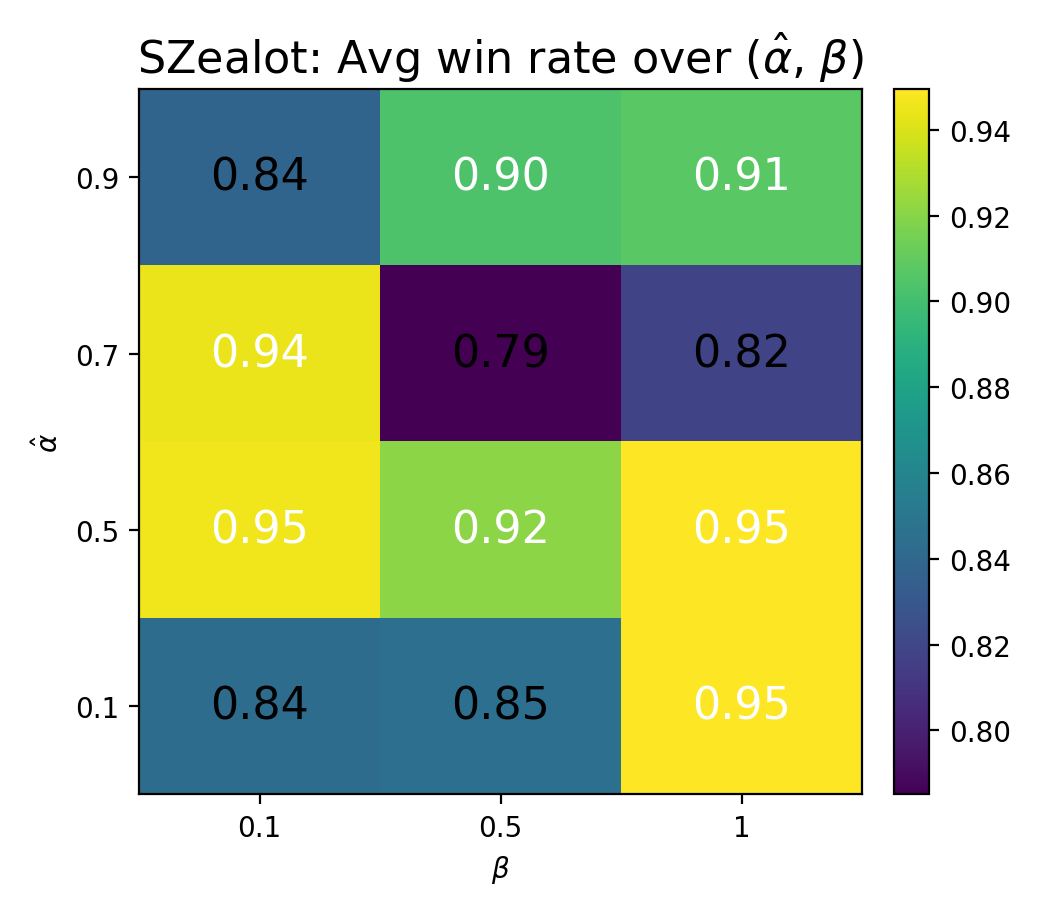}
    \caption{Stalker-Zealot}
    \label{fig:param-smac-sz}
  \end{subfigure}\hfill
  \begin{subfigure}{0.15\linewidth}
    \includegraphics[width=\linewidth]{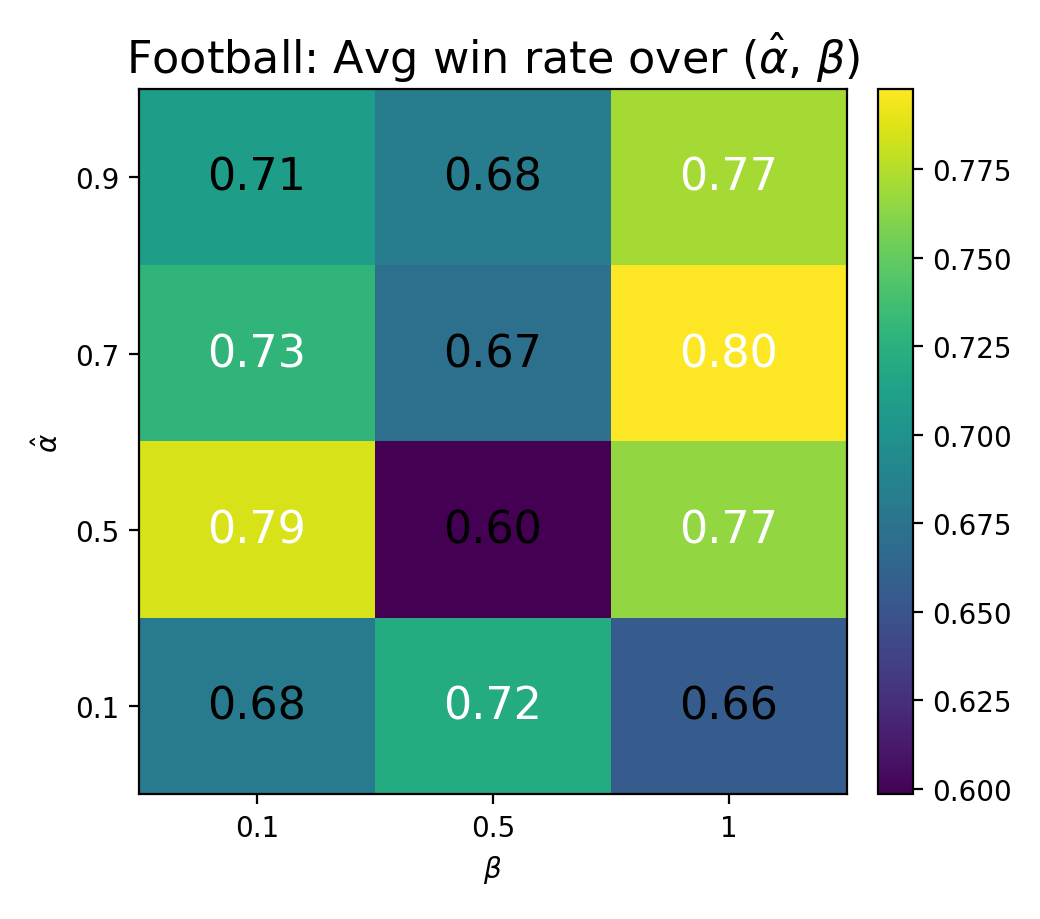}
    \caption{Football}
    \label{fig:param-football}
  \end{subfigure}

  \caption{The average win rate for different combinations of $\hat{\alpha}$ and $\beta$ used by MCS in all environments.}
  \label{fig:paramsCompare}
\end{figure*}

We also conducted a sensitivity analysis of the threshold $\hat{\alpha}$ and coefficient parameter $\beta$. We report the mean of the averaged win rate in the last 10 episodes across tasks under different values of $\hat{\alpha}$ and $\beta$ in Figure \ref{fig:paramsCompare}. In all tasks, moderate values of $\hat{\alpha}$ (e.g., 0.5 or 0.7) yield the best results, whereas larger values (e.g., 0.9) tend to degrade performance. This indicates that limited communication may negatively affect the performance, while a moderate amount of communication is sufficient to achieve good performance. The effect of $\beta$ is more subtle and task-dependent. In AliceBob and SMAC Stalker-Zealot series, both low values (e.g., $\beta=0.1$) and high values (e.g., $\beta=1$) can lead to good performance. However, in SMAC Marine and Football series, $\beta$ must be carefully tuned in combination with $\hat{\alpha}$. We also observe that inappropriate combinations of $\hat{\alpha}$ and $\beta$, for example, $\hat{\alpha}=0.9$ and $\beta=0.5$ in the AliceBob 233 series, can lead to a drastic drop in performance. This typically occurs when communication is severely limited (high $\hat{\alpha}$) or when the message encoding is overly regularized (high $\beta$). Overall, the analysis suggests that a robust configuration across most tasks is $\hat{\alpha}$=0.5 and $\beta$=0.1, suggesting moderate amount of communication and slowly regularize the messages. In fact, in the finetuning stage, we adopt a strategy of adjusting $\hat{\alpha}$ around 0.5 to regulate communication, while starting with a small value of $\beta$ and gradually increasing it.

\subsection{Analysis of Message Representations}
\label{sec:analyComm}

\begin{figure}[t]
  \centering

  \begin{subfigure}[t]{0.9\linewidth}
    \centering
    \includegraphics[width=\linewidth]{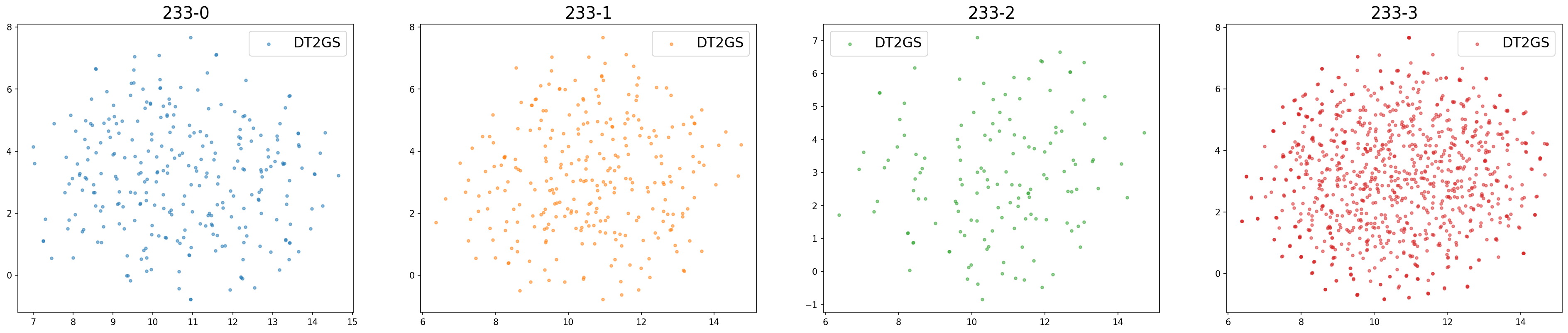}
    \subcaption{DT2GS in AliceBob}
    \label{fig:latenAliceDT2GS}
  \end{subfigure}

  \begin{subfigure}[t]{0.9\linewidth}
    \centering
    \includegraphics[width=\linewidth]{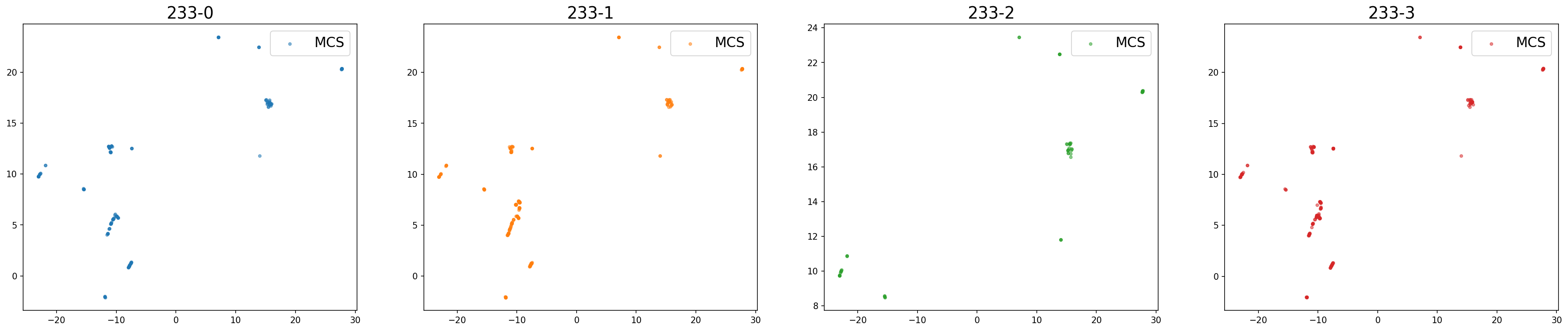}
    \subcaption{MCS in AliceBob}
    \label{fig:latenAliceMCS}
  \end{subfigure}

  \begin{subfigure}[t]{0.9\linewidth}
    \centering
    \includegraphics[width=\linewidth]{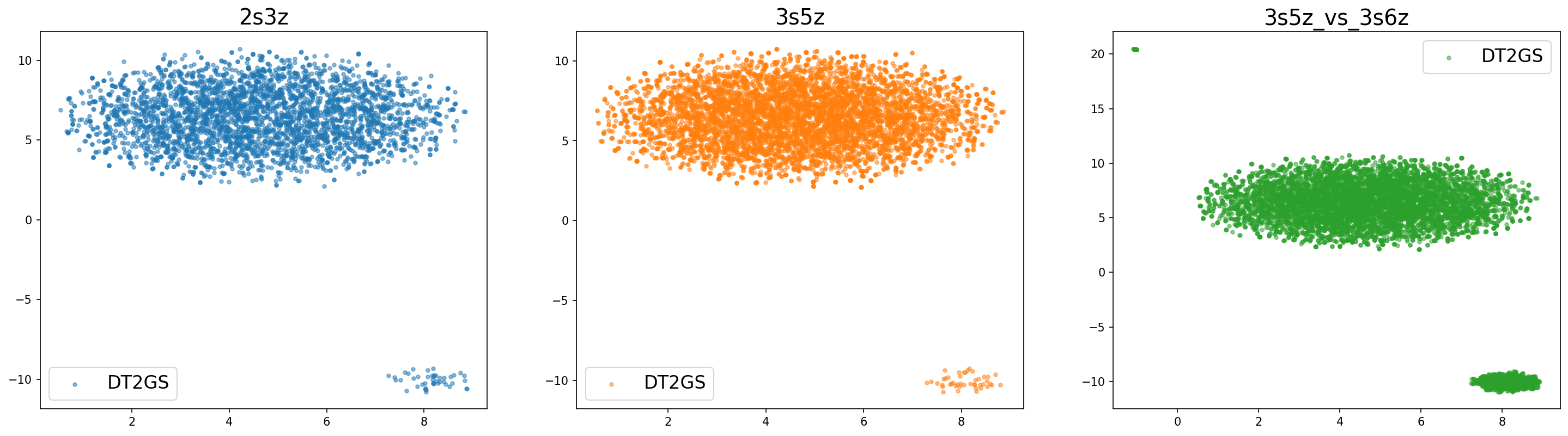}
    \subcaption{DT2GS in SMAC}
    \label{fig:latenSMACDT2GS}
  \end{subfigure}

  \begin{subfigure}[t]{0.9\linewidth}
    \centering
    \includegraphics[width=\linewidth]{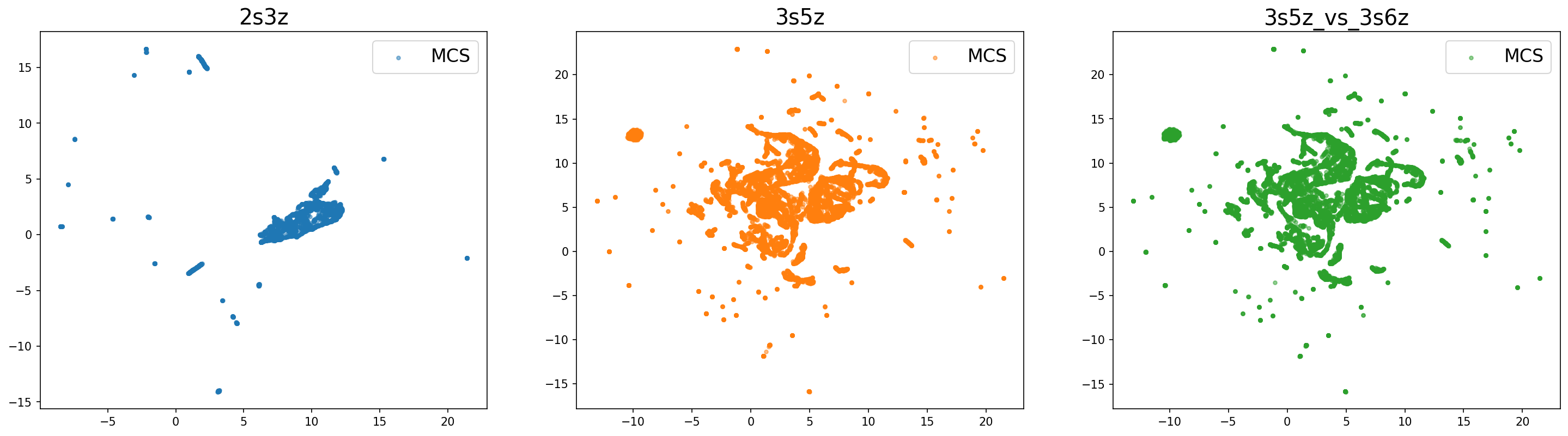}
    \subcaption{MCS in SMAC}
    \label{fig:latenSMACMCS}
  \end{subfigure}

  \begin{subfigure}[t]{0.45\linewidth}
    \centering
    \includegraphics[width=\linewidth]{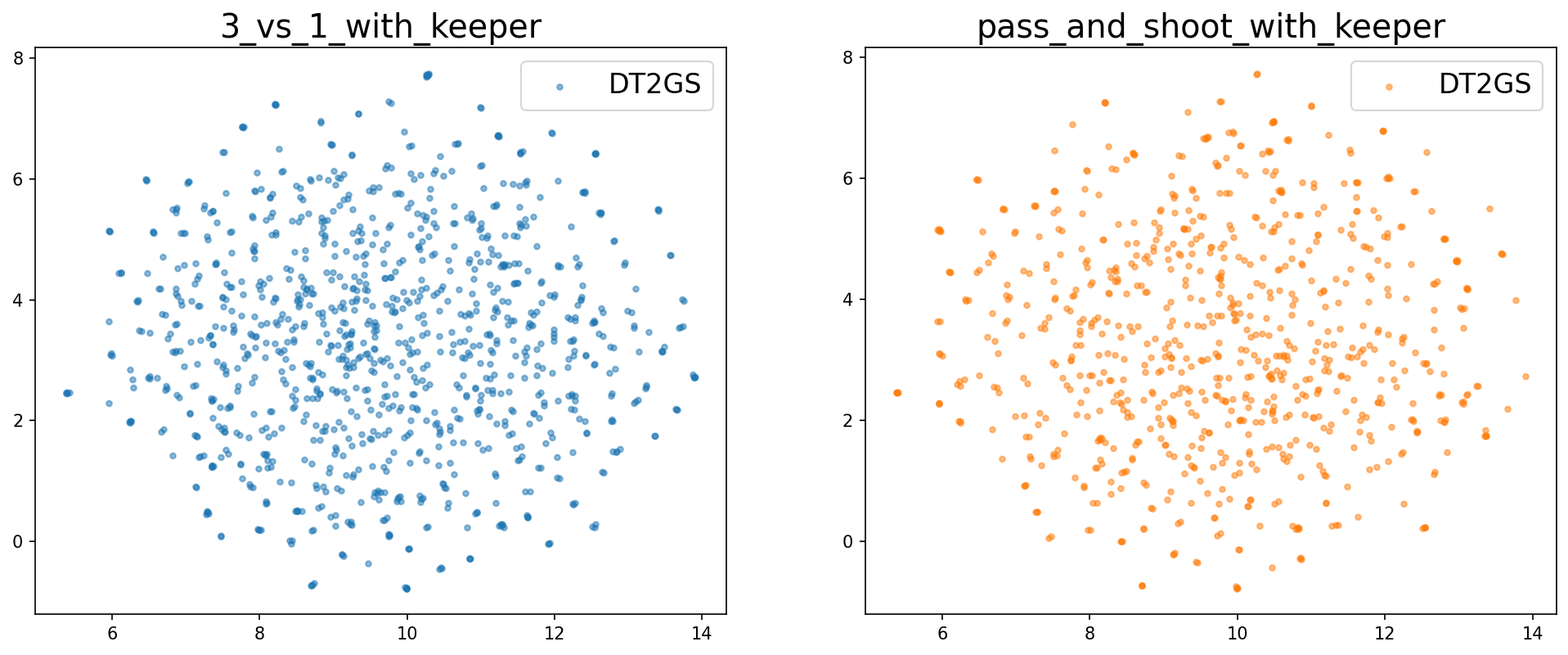}
    \subcaption{DT2GS in Football}
    \label{fig:latenFootDT2GS}
  \end{subfigure}\hfill
  \begin{subfigure}[t]{0.45\linewidth}
    \centering
    \includegraphics[width=\linewidth]{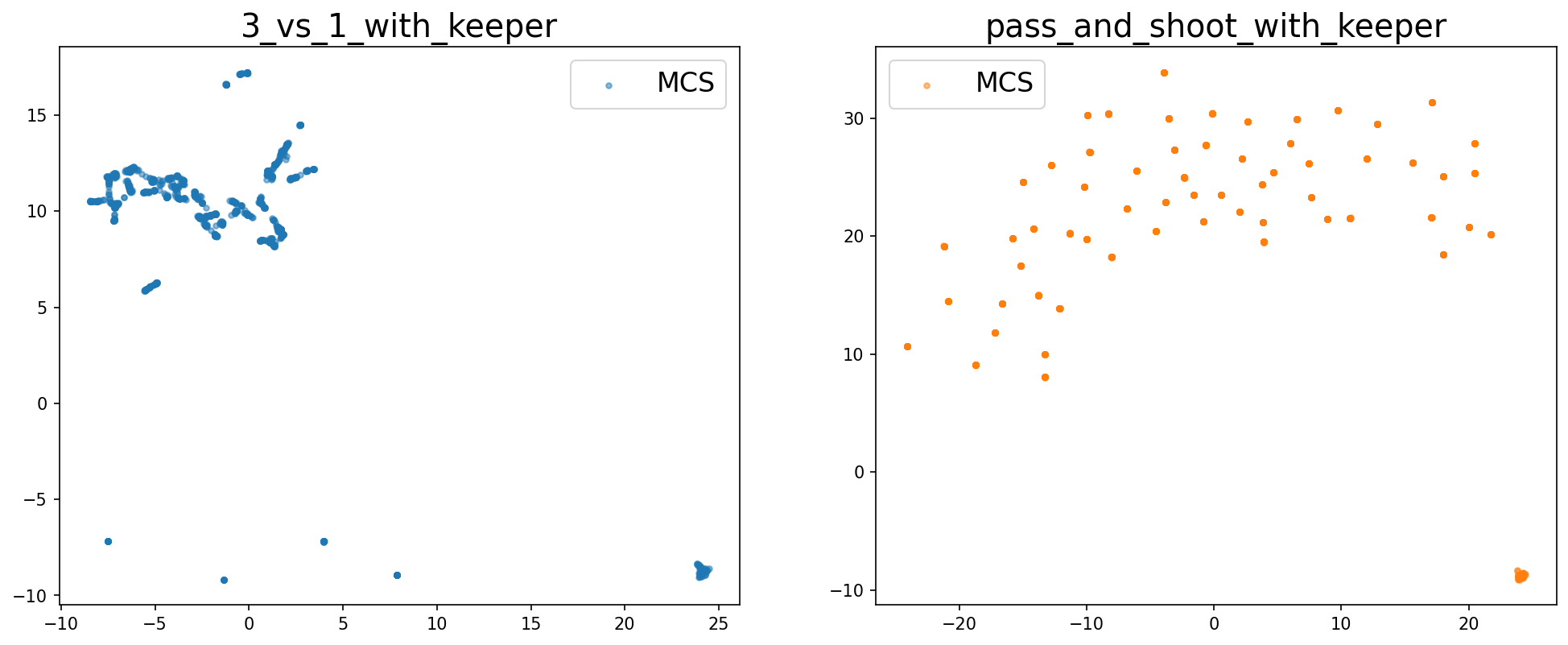}
    \subcaption{MCS in Football}
    \label{fig:latenFootMCS}
  \end{subfigure}

  \caption{Latent representations of DT2GS and MCS.}
  \label{fig:latent_combined}

\end{figure}

We further investigate the messages learned by MCS, which correspond to the latent representations generated by our proposed Transformer-based message encoder. To examine the impact of communication on learning, we compare the latent representations produced by MCS with DT2GS, which also employs a Transformer to encode observations for policy learning but does not incorporate communication. For visualization, we record the latent representations from the last three episodes of MCS and DT2GS. We then apply UMAP \cite{Leland2018UMAP} to project the high-dimensional latent representations into two dimensional points, as shown in Figure \ref{fig:latent_combined}, where each point corresponds to an agent. In AliceBob, where different tasks differ only in the types of entities, both DT2GS and MCS learn similar latent representations across tasks. However, MCS produces more compact message representations, indicating a shared communication pattern among agents, which enhances their coordination. In SMAC, DT2GS tends to learn similar representations across all tasks, even though task $2c3s$ should intuitively differ from the others due to the smaller number of agents and enemies involved. In contrast, MCS can capture this difference and learns a distinct representation for $2c3s$, likely because fewer agents participate in communication. In Football, MCS also distinguishes between $3\_vs\_1\_with\_keeper$ and $pass\_and\_shoot\_with\_keeper$, indicating that agents may adopt different shooting strategies as the number of agents and their starting positions vary across the two tasks. Furthermore, in all tasks, DT2GS tends to spread the representations in the latent space, treating agents independently without capturing their interactions. In contrast, MCS forms clusters that reflect structured representations, indicating meaningful inter-agent dependencies, which is consistent with its superior empirical performance.


\section{Conclusion}

We propose Multi-task Communication Skills (MCS), a multi-task MADRL with communication method that learns a shared communication protocol across tasks with varying numbers of agents, observation spaces, and action spaces. We introduce a prediction network that maximizes mutual information between messages and actions to promote coordinated action selection. Empirical results show that MCS outperforms multi-task MADRL methods without communication and single-task MADRL methods with or without communication across several benchmark multi-task environments. Moreover, MCS exhibits meaningful patterns in the latent message representations, while the amount of communication required may vary across different environments. In future work, we aim to develop more adaptive strategies that dynamically prune unnecessary messages for each task. We will also investigate how the differences across tasks influence the learned message representation.






\bibliographystyle{unsrt} 
\bibliography{reference}


\newpage
\appendix
\onecolumn

\section{Details of Entity-based Representations}

\begin{figure}[h] 
    \centering
    \includegraphics[width=\textwidth]{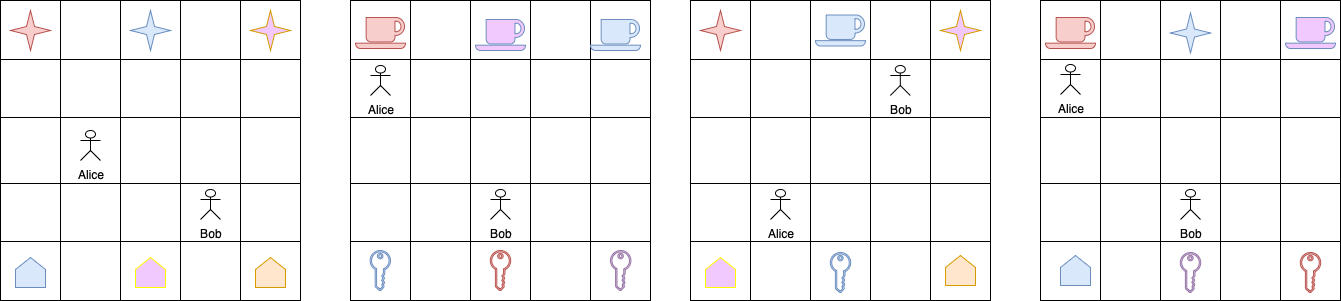}
    \caption{An illustration of the AliceBob-233 series. Agents, diamonds/food, and buttons/keys are randomly placed on the map in each episode.}
    \label{fig:alicebob}
\end{figure}

\begin{figure}[ht]
  \centering
  \begin{minipage}{0.3\textwidth}
    \centering
    \includegraphics[width=\linewidth]{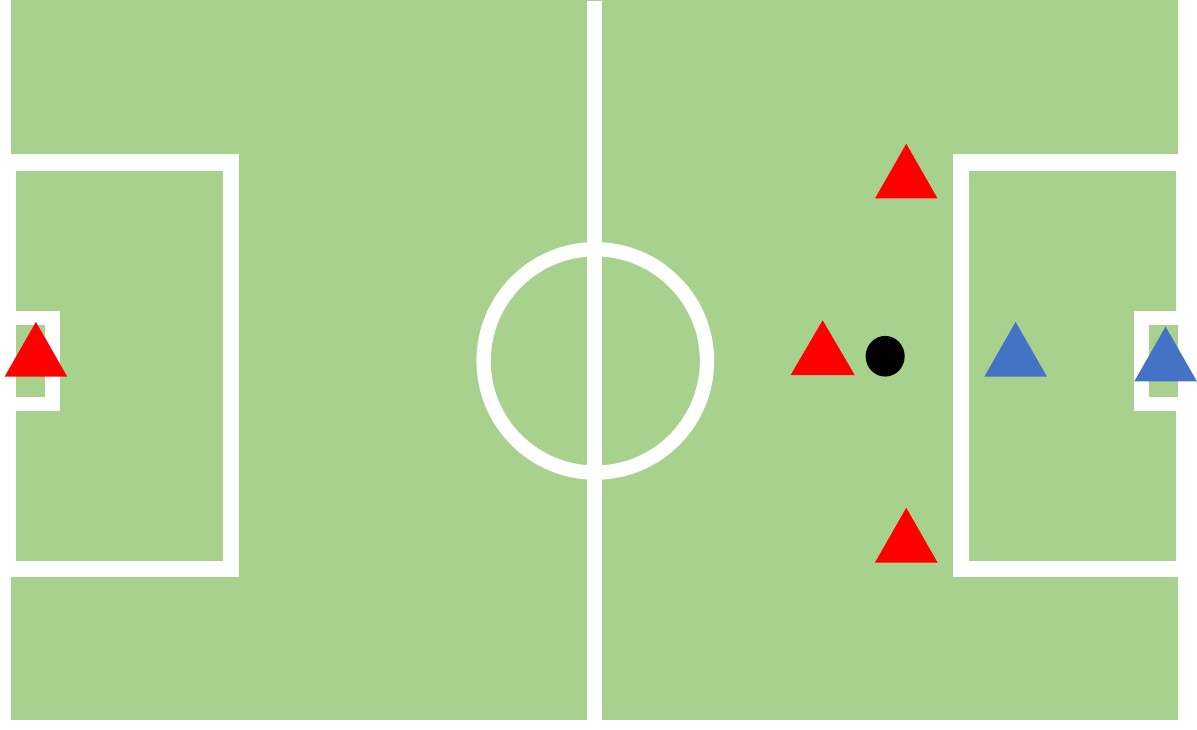}
  \end{minipage}
  \begin{minipage}{0.3\textwidth}
    \centering
    \includegraphics[width=\linewidth]{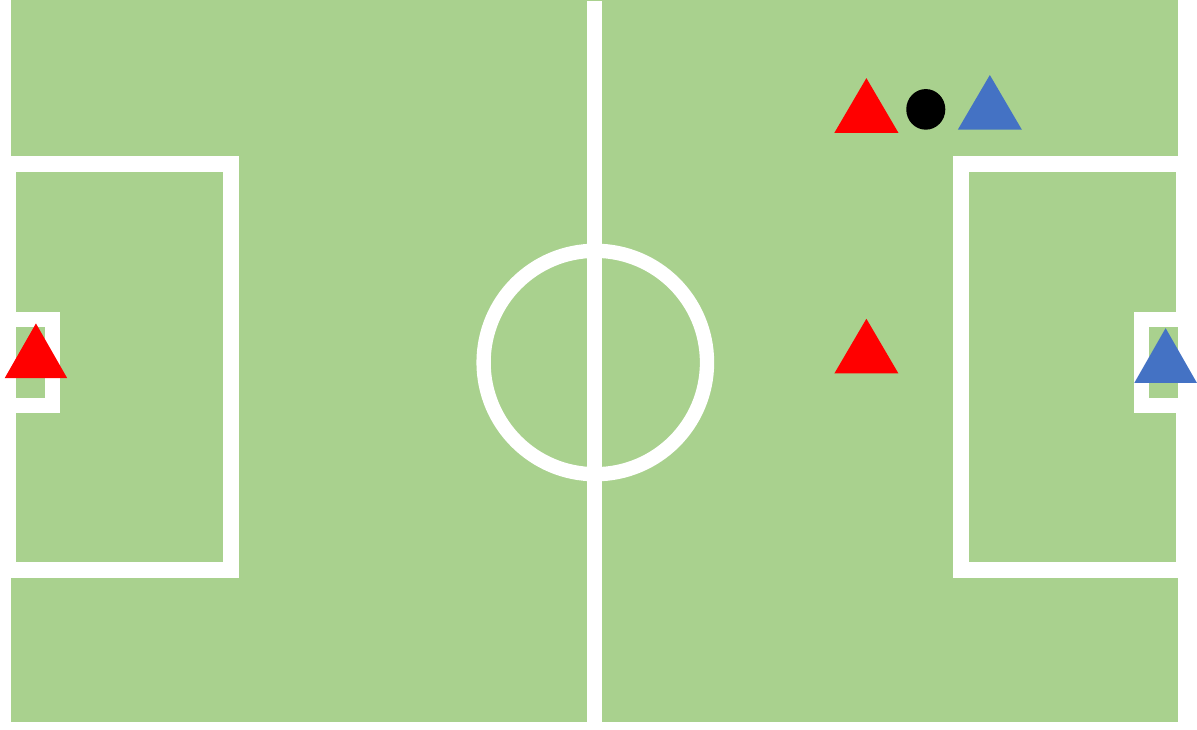}
  \end{minipage}
  \caption{Football series with 3 vs 1 with keeper (Left) and Pass and shot with keeper (Right).}
  \label{fig:my_figure}
\end{figure}

\textbf{AliceBob series}. The AliceBob environment was originally proposed for single-task MADRL without considering entity-based representations. Following the approaches of Hu et al. \cite{Hu2021UPDet} and Tian et al. \cite{Tian2024DT2GS}, we construct an entity-based representation and adapt the AliceBob environment to multi-task settings. In the entity-based representation, entities consist of RL agents, buttons/keys, and diamonds/food. For each entity, we extract partially observable features, including the relative distance from each agent to all other entities and a one-hot representation of the observed entity type. To reduce learning complexity, we further enrich each agent’s observation by incorporating features of the surrounding grid cells using the same representation scheme. The resulting entity representation thus comprises: (i) features of the agent itself, (ii) features of other agents, (iii) features of diamonds/food, (iv) features of keys/buttons, and (v) features of the surrounding grids. For example, in the AliceBob-233 series, the observation shape becomes $16 \times 17$ where 16 is the total number of entities (8 original entities plus 8 surrounding grids), and 17 is the number of features (2 dimensions for the relative distance and 15 dimensions for the one-hot encoding). We provide an example of the multi-task setting in the AliceBob-233 series in Figure \ref{fig:alicebob}, where two agents must coordinate to collect diamonds/food and press buttons/keys separately. The type of entities is determined before the start of the game, while their positions are randomly generated at the beginning of each episode to create dynamic targets. Agents can leverage the knowledge acquired from collecting diamonds or pressing buttons in one task to improve their learning performance in another task.

\textbf{SMAC series}. We use the same multi-task environment as used by Hu et al. \cite{Hu2021UPDet} and Tian et al. \cite{Tian2024DT2GS}, where entities consist of the agent itself, allied agents, and enemies. The feature set includes visibility, distance to the agent, relative $x$ and $y$ positions, health, shield, and unit type of each entity. Then, for each entity, a total of 16 feature dimensions are considered \footnote{Implementation available at: \url{https://github.com/Felixvillas/DT2GS}}.

\textbf{Football series}. We adapt the Google Research Football environment \cite{Kurach2020Football} to a multi-task setting with entity-based representations. Starting from the raw features in \texttt{simple115v2}, we group them into four macro-entities: left-team locations, left-team directions, right-team locations, and right-team directions. Ball position and direction are appended to the appropriate entity depending on possession, and a one-hot encoding of the active player is included. This results in 43 feature dimensions per entity, with the total number of features across all entities remaining similar to the original raw representation, thereby preserving the overall complexity of the environment. We illustrate two tasks from the Football series $\{3\_vs\_1\_with\_keeper, pass\_and\_shoot\_with\_keeper\}$. In $3\_vs\_1\_with\_keeper$, three left-team players start in the right half, competing against one right-team defender and the goalkeeper. In $pass\_and\_shoot\_with\_keeper$, two left-team players start in the right half against one right-team defender and the goalkeeper. An episode terminates when (a) the maximum duration (200 steps) is reached, (b) the ball goes out of bounds, (c) a team scores, or (d) possession changes.

\section{Critical Hyperarameters and Per-task Performance}

\begin{table}[t]
    \caption{\textcolor{black}{Default hyperparameters of MCS across different environments.}}
    \label{tab:mcs_hparams}
    \begin{center}
    \begin{small}
    \begin{sc}
    \begin{tabular}{lccc}
        \toprule
        \textbf{Hyperparameter} & \textbf{AliceBob} & \textbf{SMAC} & \textbf{Football} \\
        \midrule
        $lr$                         & 5e-4  & 5e-4  & 5e-4  \\
        $critic\_lr$                & 5e-4  & 5e-4  & 5e-4  \\
        $opti\_eps$                 & 1e-5  & 1e-5  & 1e-5  \\
        PPO epochs                  & 8    & 8    & 15    \\
        mini-batches                & 10     & 8     & 2     \\
        $clip\_param$              & 0.2   & 0.2   & 0.2   \\
        entropy coefficient         & 0.01  & 0.01  & 0.01  \\
        max grad norm               & 10    & 10    & 10    \\
        $\gamma$                    & 0.99  & 0.99  & 0.99  \\
        GAE $\lambda$              & 0.95  & 0.95  & 0.95  \\
        hidden size                 & 64    & 64    & 64    \\
        message dimension           & 10    & 10    & 10    \\
        batch size                  & 32    & 32    & 32    \\
        number of steps              & 1e6   & 5e6   & 1e7   \\
        evaluation episodes         & 32    & 32    & 32    \\        
        \bottomrule
    \end{tabular}
    \end{sc}
    \end{small}
    \end{center}
\end{table}

\begin{table*}[t]
\centering
\caption{Hyperparameters (\(\hat{\alpha}\) threshold and \(\beta\)) used by MCS in different multi-task series.}
\label{tab:mcs_params}
\begin{tabular}{lcccccc}
\toprule
\textbf{Parameters} & AliceBob 233 & AliceBob 344 & SMAC Marine & SMAC Stalker Zealots & Football \\
\midrule
\(\hat{\alpha}\) & 0.5 & 0.7 & 0.9 & 0.5 & 0.5 \\
\(\beta\) & 0.1 & 1.0 & 0.1 & 0.5 & 0.1 \\
\bottomrule
\end{tabular}
\end{table*}

The hyperparameters used for MCS in AliceBob, SMAC, and Google Research Football are summarized in Table \ref{tab:mcs_hparams}. As shown, the three environments share most hyperparameters, except for the number of PPO epochs and the mini-batch size. For SMAC, we use the same hyperparameters as in DT2GS, which have already been fine-tuned. For AliceBob, we adopt similar values but use a slightly larger mini-batch size to obtain more samples. For the Football environment, we use the same hyperparameters as those employed in the single-task setting. Consequently, these hyperparameters are also applied to the baseline methods (e.g., DT2GS and RIT) to ensure a fair comparison. We also report the communication threshold ($\hat{\alpha}$) and coefficient parameter ($\beta$) used by MCS in Table \ref{tab:mcs_params}, which correspond to the results presented in Figures 4 and 5 of the manuscript.

\begin{table}[t]
  \centering
  \caption{Average wall-clock runtime (\textbf{hours}) per method across tasks under each multi-task series.}
  \label{tab:runtime_summary_no_avg_int}
  \begin{tabular}{lrrrrr}
    \toprule
    \textbf{Method} & \textbf{AliceBob 233 series} & \textbf{AliceBob 344 series} & \textbf{SMAC Marines} & \textbf{SMAC Stalker Zealot} & \textbf{Football} \\
    \midrule
    MCS    & 4 & 4 & 23 & 10 & 13 \\
    DT2GS  & 6 & 6 & 7  & 9  & 10 \\
    RIT    & 7 & 8 & 67 & 81 & 116 \\
    HMASD  & 1 & 1 & 6  & 6  & 8 \\
    MAT    & 1 & 1 & 6  & 8  & 8 \\
    MAIC   & 4 & 4 & 6  & 9  & 29 \\
    TGCNet & 4 & 4 & 13 & 12 & 25 \\
    \bottomrule
  \end{tabular}
\end{table}

The experiments were run in parallel on a cluster with 32 CPU cores and an NVIDIA A100 GPU. Table \ref{tab:runtime_summary_no_avg_int} reports the wall-clock training time for each method on each multi-task series; values are averaged over the tasks within a series. On AliceBob, the single-task methods HMASD and MAT are most compute-efficient, typically finishing within $\leq 1$ hour per run. DT2GS is moderately heavier ($6$h), while MCS/MAIC/TGCNet sit around $4\text{–}5$h. RIT is the slowest even on small tasks ($7\text{–}8$h). On SMAC (Marine and Stalker Zealot) and Football, runtimes spread more widely. RIT scales poorly (tens to $>100$h), suggesting unfavorable difficulty in complex environments. MCS remains reasonable on simpler tasks but is longer on Marines ($23$h). MAIC is fast on Marines ($6$h) yet slow on Football ($29$h). Overall, the proposed MCS does not significantly increase runtime compared to other multi-task MADRL baselines on AliceBob and Football, though it is slightly slower than DT2GS on Marines. Single-task baselines (HMASD/MAT) achieve much lower per-series runtime as they train only one task at a time, which may come at the expense of multi-task learning performance.

\begin{figure}[t]
  \centering

  \begin{subfigure}{\columnwidth}
    \centering
    \includegraphics[width=0.23\columnwidth]{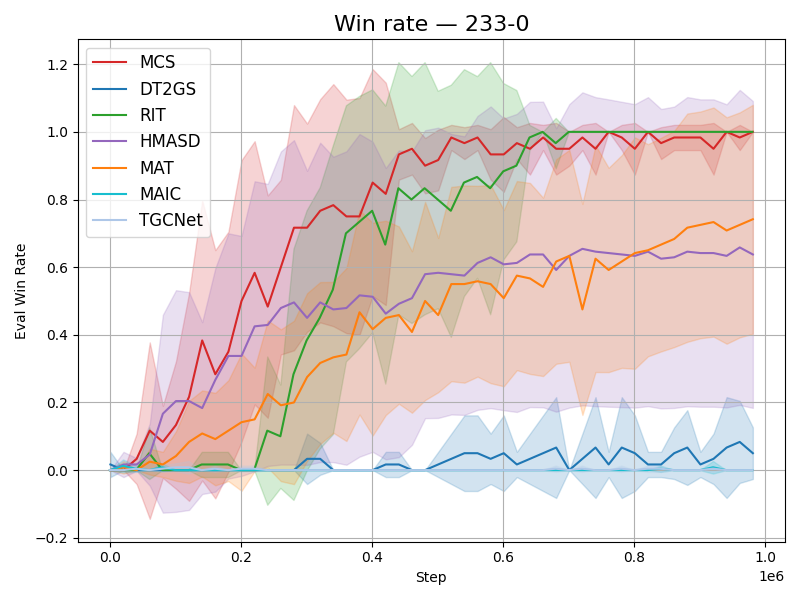}
    \includegraphics[width=0.23\columnwidth]{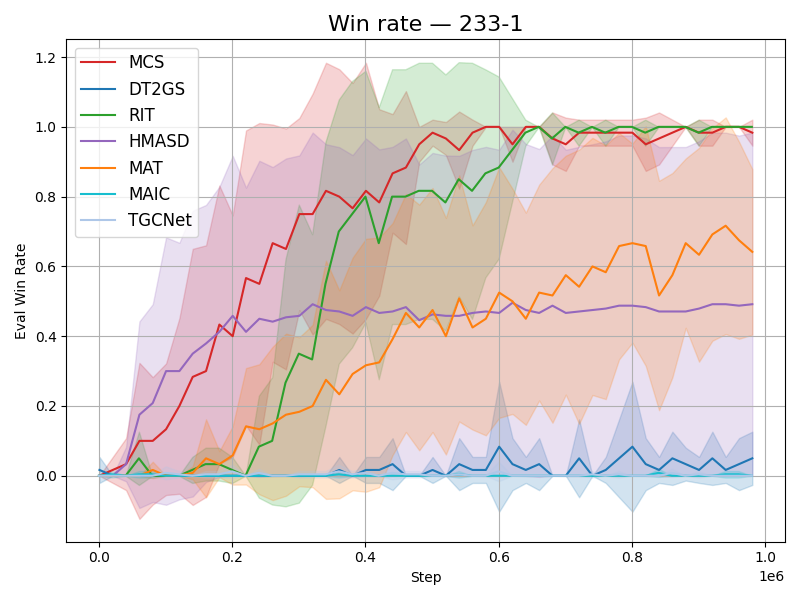}
    \includegraphics[width=0.23\columnwidth]{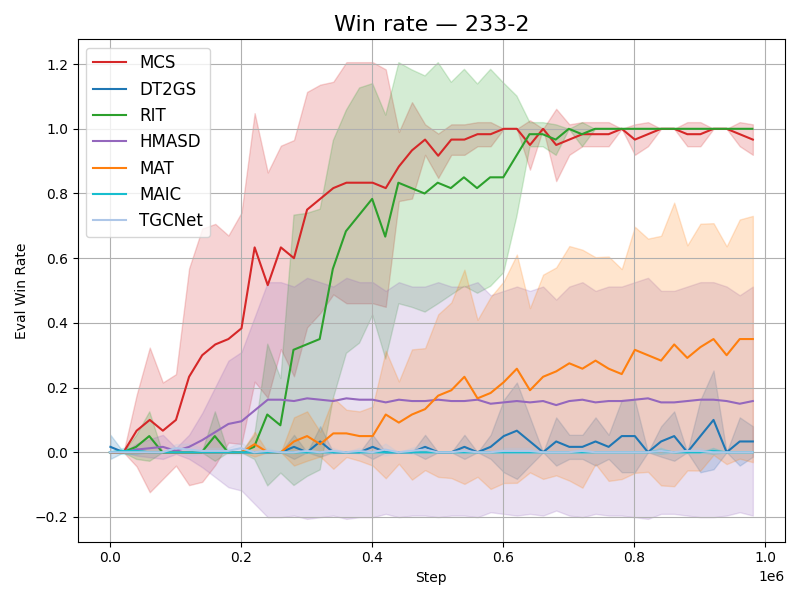}
    \includegraphics[width=0.23\columnwidth]{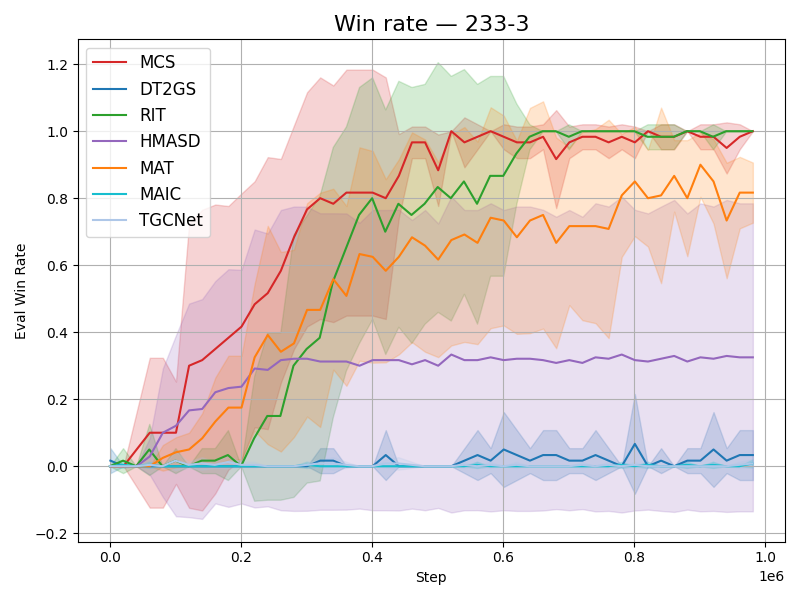}
    \caption{AliceBob-233}
    \label{fig:alicebob233}
  \end{subfigure}

  \begin{subfigure}{\columnwidth}
    \centering
    \includegraphics[width=0.23\columnwidth]{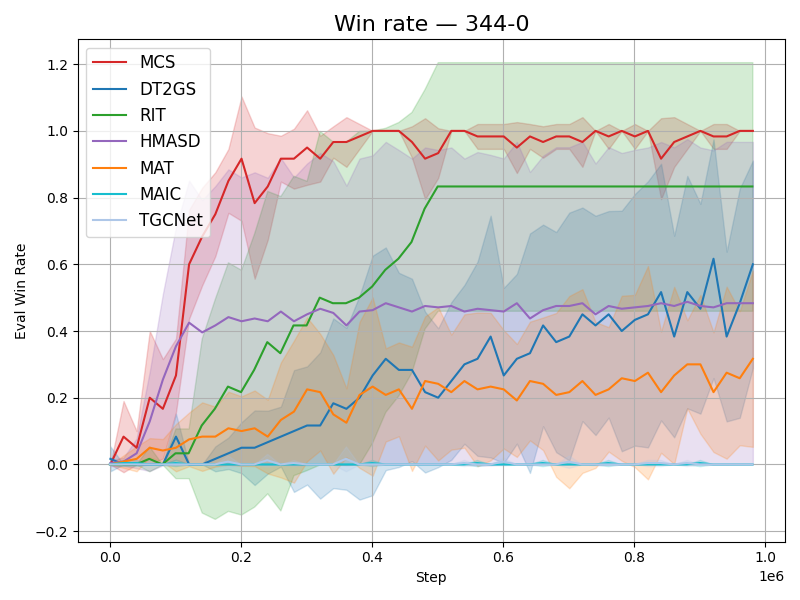}
    \includegraphics[width=0.23\columnwidth]{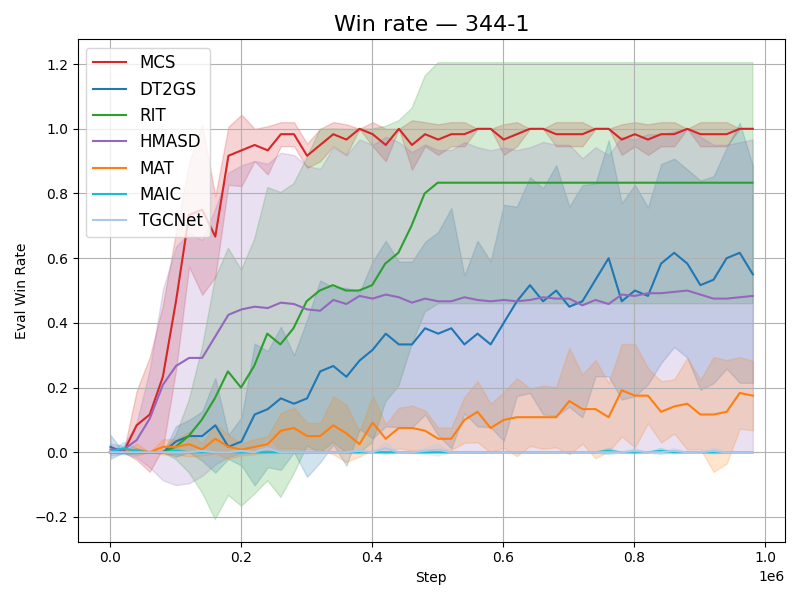}
    \includegraphics[width=0.23\columnwidth]{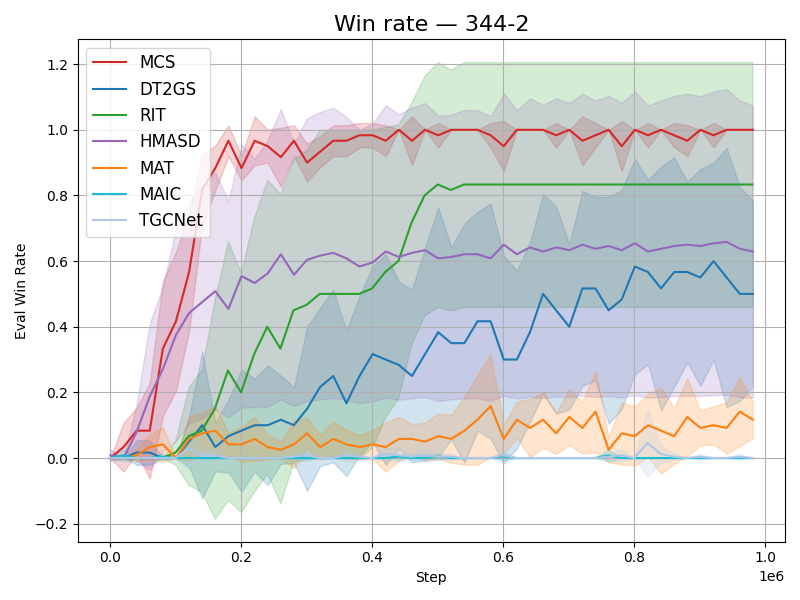}
    \includegraphics[width=0.23\columnwidth]{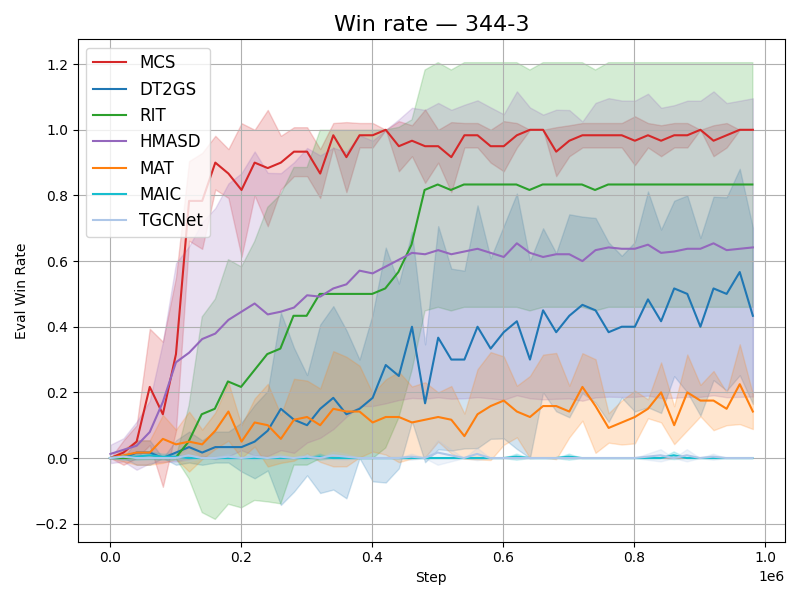}
    \subcaption{AliceBob-344}
    \label{fig:alicebob344}
  \end{subfigure}

  \begin{subfigure}{\columnwidth}
    \centering
    \includegraphics[width=0.23\columnwidth]{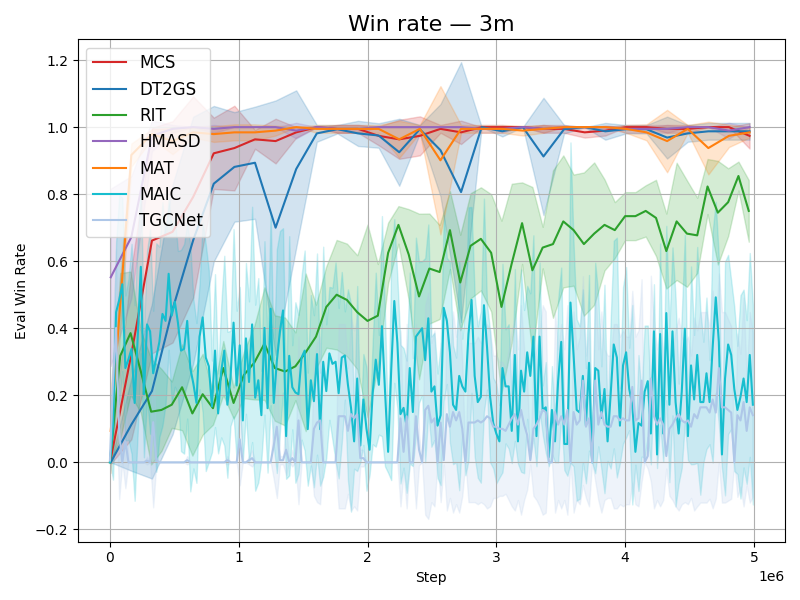}
    \includegraphics[width=0.23\columnwidth]{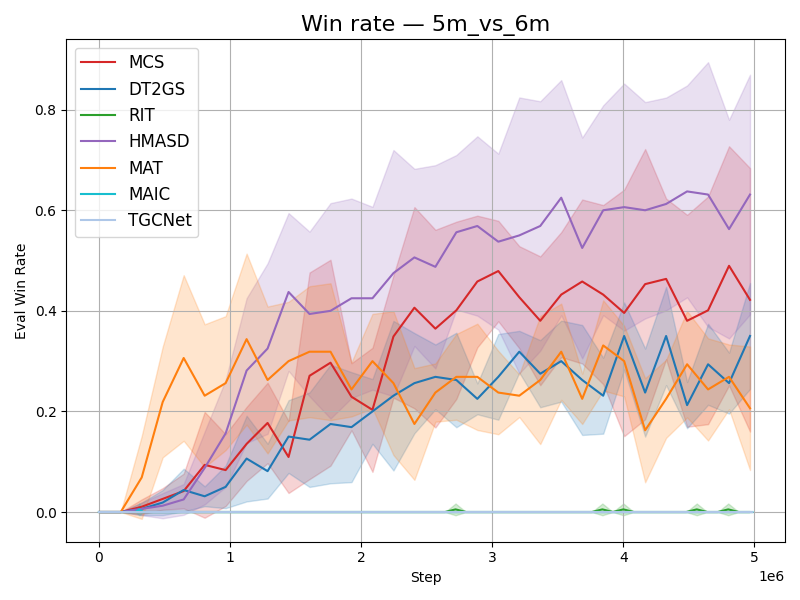}
    \includegraphics[width=0.23\columnwidth]{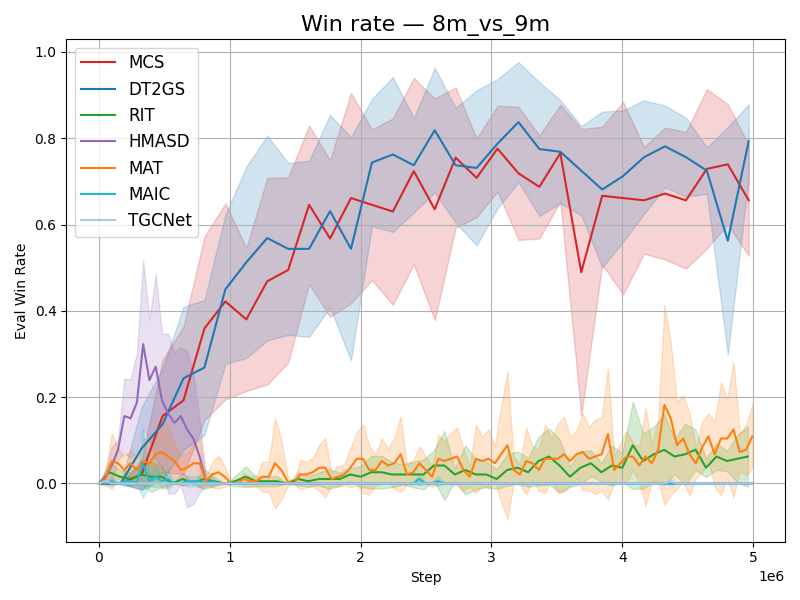}
    \includegraphics[width=0.23\columnwidth]{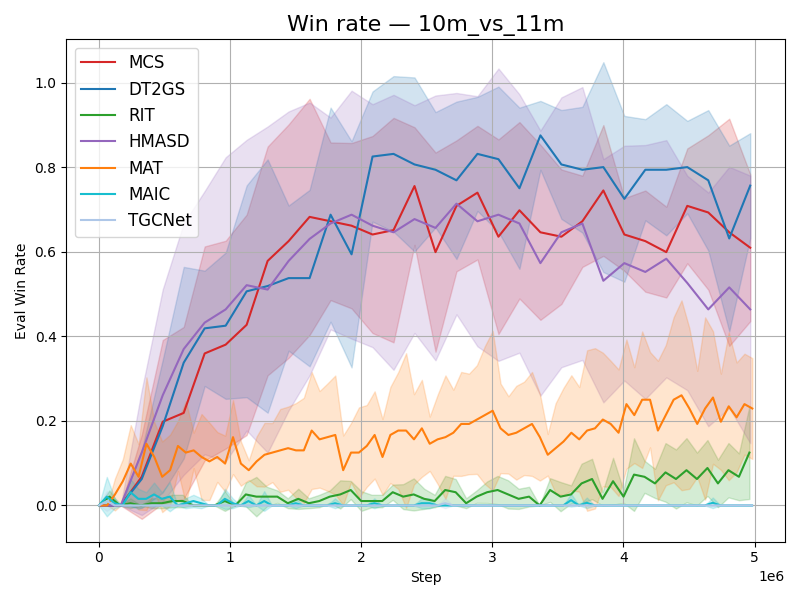}
    \caption{Marine}
    \label{fig:Marine}
  \end{subfigure}

  \begin{subfigure}{\columnwidth}
    \centering
    \includegraphics[width=0.23\columnwidth]{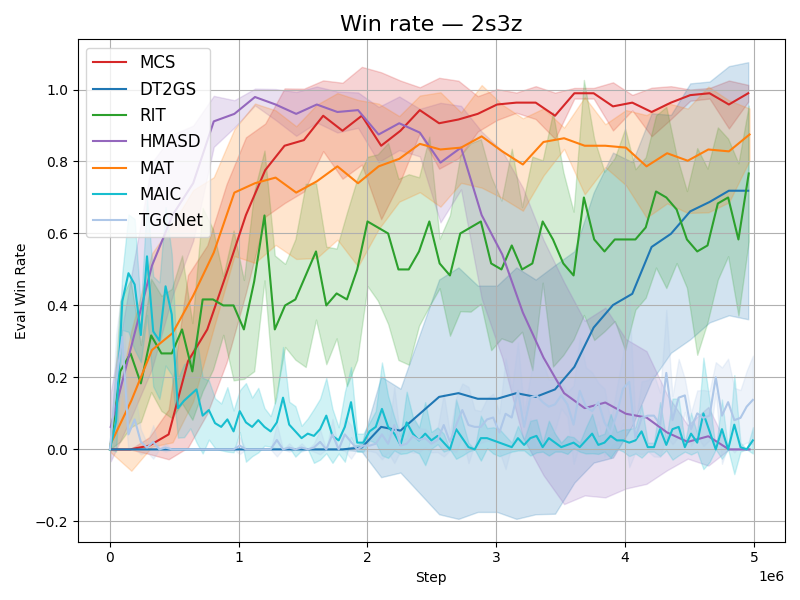}
    \includegraphics[width=0.23\columnwidth]{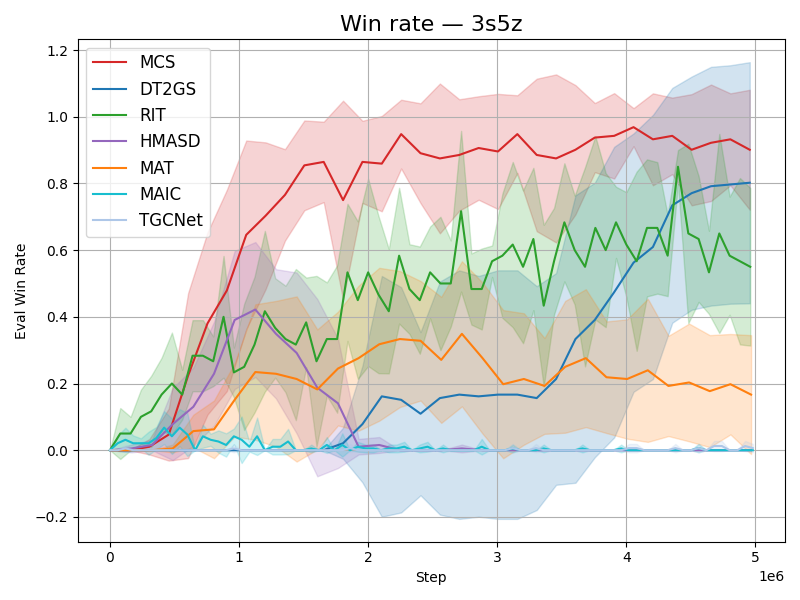}
    \includegraphics[width=0.23\columnwidth]{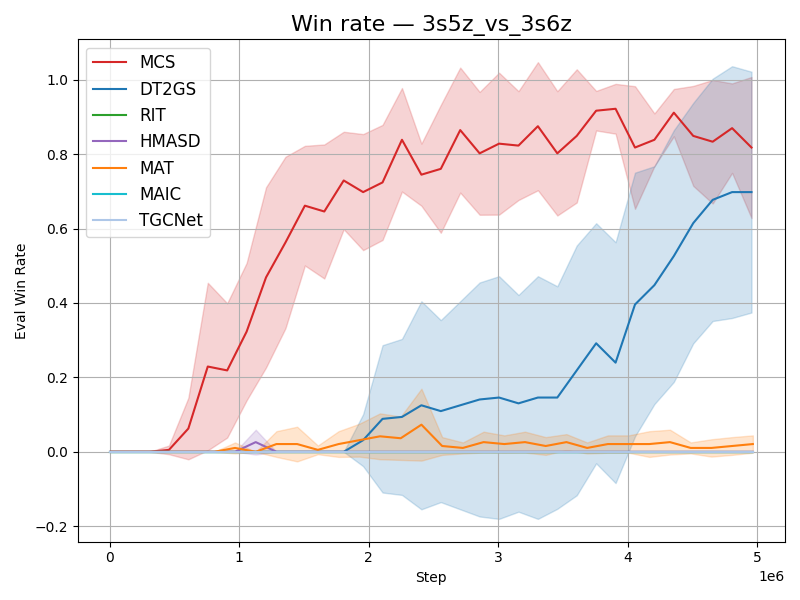}
    \caption{Stalker Zealots}
    \label{fig:SZealots}
  \end{subfigure}

  \begin{subfigure}{\columnwidth}
    \centering
    \includegraphics[width=0.3\columnwidth]{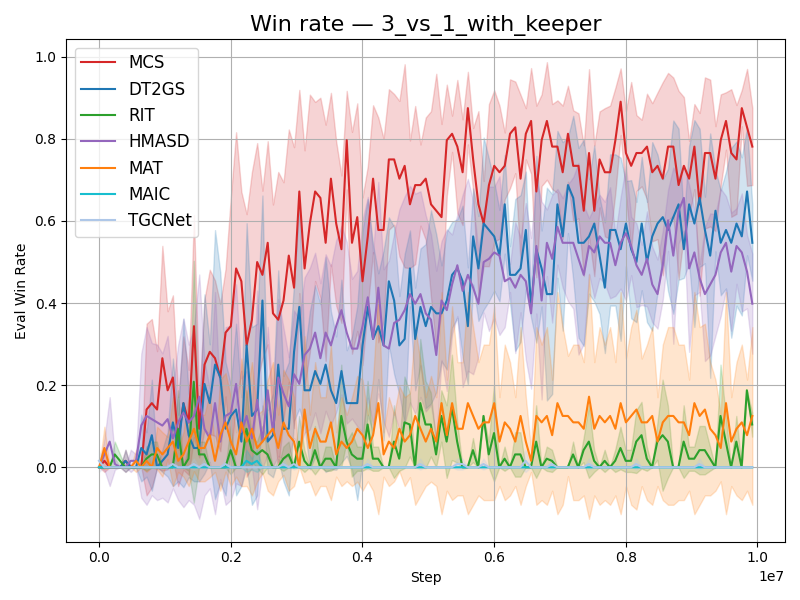}
    \includegraphics[width=0.3\columnwidth]{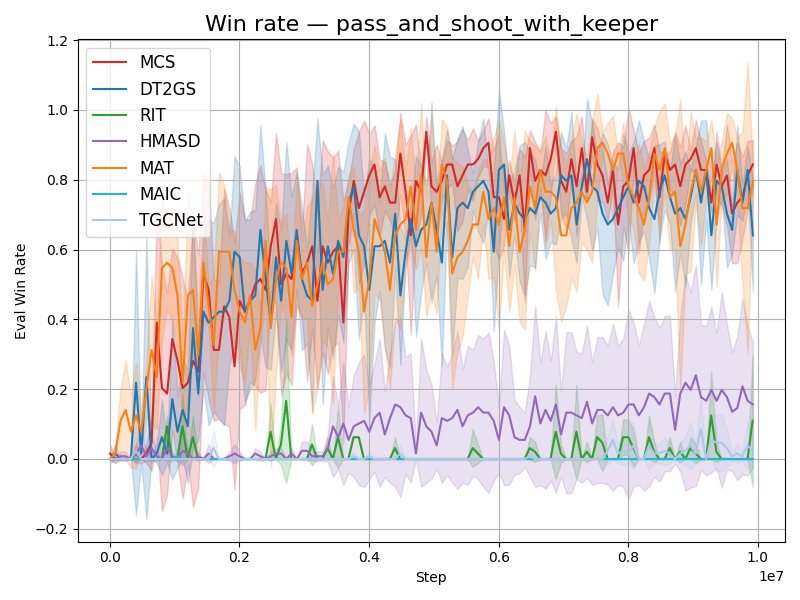}
    \caption{Football}
    \label{fig:Football}
  \end{subfigure}

  \caption{Per-task win rate curves across different environments: AliceBob-233/344, SMAC Marine/Stalker Zealots, and Football.}
  \label{fig:per_winrate}
\end{figure}

We further report the per-task performance of each method across all multi-task series, as shown in Figure \ref{fig:per_winrate} and Table \ref{tab:grouped_winrates_with_avgs}. As illustrated in Figures \ref{fig:alicebob233}–\ref{fig:alicebob344}, MCS consistently achieves faster convergence and significantly higher win rates across all tasks in the AliceBob environments. In SMAC Marines (Figure \ref{fig:Marine}), MCS outperforms DT2GS in the $3m$ and $5m\_vs\_6m$ tasks, while DT2GS surpasses MCS in $8m\_vs\_9m$ and $10m\_vs\_11m$, potentially due to the benefits of task decomposition in the two tasks. Despite these differences, MCS and DT2GS achieve similar average performance across tasks (as shown in the table). For single-task baselines, methods such as HMASD achieve high win rates in $5m\_vs\_6m$ and $10m\_vs\_11m$. However, their performance varies significantly across tasks, resulting in lower overall average performance. In the Stalker Zealots series, MCS consistently outperforms all baselines in terms of learning performance. In Football, MCS achieves a significantly higher win rate in the $3\_vs\_1\_with\_keeper$ task. In the $pass\_and\_shoot\_with\_keeper$ task, the single-task baseline MAT performs similar to MCS. However, MAT’s performance drops significantly in $3\_vs\_1\_with\_keeper$, where MCS maintains a consistent high win rate.

\begin{table}[t]
  \centering
  \scriptsize
  \caption{Per-task win rates ($mean(std)$) for all benchmark multi-task series. Columns list individual tasks; the rightmost column is the average across tasks.}
  \label{tab:grouped_winrates_with_avgs}
  \resizebox{\columnwidth}{!}{%
  \begin{tabular}{lccccc}
    \toprule
    \multicolumn{6}{c}{\textbf{AliceBob 233 series}} \\
    \cmidrule(lr){1-6}
    \textbf{Method} & \textbf{233-0} & \textbf{233-1} & \textbf{233-2} & \textbf{233-3} & \textbf{Avg} \\
    \midrule
    MCS    & 0.98 (0.02) & 0.98 (0.02) & 0.99 (0.01) & 0.98 (0.01) & \textbf{0.98 (0.00)} \\
    DT2GS  & 0.05 (0.09) & 0.04 (0.08) & 0.04 (0.06) & 0.02 (0.06) & 0.04 (0.01) \\
    RIT    & 1.00 (0.00) & 1.00 (0.01) & 1.00 (0.00) & 0.99 (0.01) & \textbf{1.00 (0.00)} \\
    HMASD  & 0.64 (0.50) & 0.48 (0.53) & 0.16 (0.39) & 0.32 (0.50) & 0.40 (0.21) \\
    MAT    & 0.70 (0.35) & 0.64 (0.30) & 0.32 (0.40) & 0.82 (0.13) & 0.62 (0.21) \\
    MAIC   & 0.00 (0.00) & 0.00 (0.00) & 0.00 (0.00) & 0.00 (0.00) & 0.00 (0.00) \\
    TGCNet & 0.00 (0.00) & 0.00 (0.00) & 0.00 (0.00) & 0.00 (0.00) & 0.00 (0.00) \\
    \midrule
    \multicolumn{6}{c}{\textbf{AliceBob 344 series}} \\
    \cmidrule(lr){1-6}
    \textbf{Method} & \textbf{344-0} & \textbf{344-1} & \textbf{344-2} & \textbf{344-3} & \textbf{Avg} \\
    \midrule
    MCS    & 0.98 (0.02) & 0.99 (0.01) & 0.99 (0.01) & 0.98 (0.01) & \textbf{0.99 (0.00)} \\
    DT2GS  & 0.49 (0.35) & 0.56 (0.33) & 0.55 (0.32) & 0.47 (0.28) & 0.52 (0.04) \\
    RIT    & 0.83 (0.41) & 0.83 (0.41) & 0.83 (0.41) & 0.83 (0.41) & 0.83 (0.00) \\
    HMASD  & 0.48 (0.52) & 0.49 (0.53) & 0.64 (0.50) & 0.64 (0.49) & 0.56 (0.09) \\
    MAT    & 0.27 (0.24) & 0.15 (0.11) & 0.10 (0.06) & 0.16 (0.05) & 0.17 (0.07) \\
    MAIC   & 0.00 (0.00) & 0.00 (0.00) & 0.00 (0.00) & 0.00 (0.00) & 0.00 (0.00) \\
    TGCNet & 0.00 (0.00) & 0.00 (0.00) & 0.01 (0.01) & 0.00 (0.01) & 0.00 (0.00) \\
    \midrule
    \multicolumn{6}{c}{\textbf{SMAC Marine series}} \\
    \cmidrule(lr){1-6}
    \textbf{Method} & \textbf{3m} & \textbf{5m\_vs\_6m} & \textbf{8m\_vs\_9m} & \textbf{10m\_vs\_11m} & \textbf{Avg} \\
    \midrule
    MCS    & 0.99 (0.00) & 0.43 (0.20) & 0.67 (0.11) & 0.66 (0.08) & \textbf{0.69 (0.23)} \\
    DT2GS  & 0.99 (0.02) & 0.28 (0.03) & 0.73 (0.12) & 0.77 (0.16) & \textbf{0.69 (0.29)} \\
    RIT    & 0.74 (0.07) & 0.00 (0.00) & 0.06 (0.04) & 0.08 (0.05) & 0.22 (0.35) \\
    HMASD  & 1.00 (0.00) & 0.60 (0.24) & 0.00 (0.00) & 0.55 (0.31) & 0.54 (0.41) \\
    MAT    & 0.98 (0.02) & 0.26 (0.06) & 0.09 (0.07) & 0.23 (0.14) & 0.39 (0.40) \\
    MAIC   & 0.24 (0.23) & 0.00 (0.00) & 0.00 (0.00) & 0.00 (0.00) & 0.06 (0.12) \\
    TGCNet & 0.13 (0.26) & 0.00 (0.00) & 0.00 (0.00) & 0.00 (0.00) & 0.03 (0.07) \\
    \midrule
    \multicolumn{5}{c}{\textbf{SMAC Stalker Zealots series}} & \\
    \cmidrule(lr){1-5}
    \textbf{Method} & \textbf{2s3z} & \textbf{3s5z} & \textbf{3s5z\_vs\_3s6z} & \textbf{Avg} & \\
    \midrule
    MCS    & 0.97 (0.02) & 0.93 (0.15) & 0.86 (0.12) & \textbf{0.92 (0.06)} & \\
    DT2GS  & 0.53 (0.33) & 0.63 (0.36) & 0.48 (0.31) & 0.55 (0.07) & \\
    RIT    & 0.65 (0.19) & 0.63 (0.16) & 0.00 (0.00) & 0.43 (0.37) & \\
    HMASD  & 0.07 (0.16) & 0.00 (0.00) & 0.00 (0.00) & 0.02 (0.04) & \\
    MAT    & 0.83 (0.11) & 0.21 (0.18) & 0.02 (0.01) & 0.36 (0.43) & \\
    MAIC   & 0.03 (0.06) & 0.00 (0.00) & 0.00 (0.00) & 0.01 (0.02) & \\
    TGCNet & 0.11 (0.07) & 0.00 (0.00) & 0.00 (0.00) & 0.04 (0.06) & \\
    \midrule
    \multicolumn{4}{c}{\textbf{Football series}} & \\
    \cmidrule(lr){1-4}
    \textbf{Method} & \textbf{3\_vs\_1\_with\_keeper} & \textbf{pass\_and\_shoot\_with\_keeper} & \textbf{Avg} & \\
    \midrule
    MCS    & 0.79 (0.12) & 0.78 (0.07) & \textbf{0.79 (0.00)} & \\
    DT2GS  & 0.58 (0.15) & 0.74 (0.08) & 0.66 (0.12) & \\
    RIT    & 0.06 (0.03) & 0.04 (0.04) & 0.05 (0.02) & \\
    HMASD  & 0.48 (0.12) & 0.17 (0.26) & 0.33 (0.22) & \\
    MAT    & 0.10 (0.19) & 0.81 (0.05) & 0.45 (0.50) & \\
    MAIC   & 0.00 (0.00) & 0.00 (0.00) & 0.00 (0.00) & \\
    TGCNet & 0.00 (0.00) & 0.03 (0.05) & 0.01 (0.02) & \\
    \bottomrule
  \end{tabular}
  }
\end{table}

\end{document}